\documentclass[aps,showpacs,prb,floatfix]{revtex4}%
\usepackage{amsfonts}
\usepackage{amsmath}
\usepackage{amssymb}
\usepackage{graphicx}%
\begin{document}
\title{A Quantum Non-demolition measurement of Fock states of mesoscopic mechanical oscillators}
\author{D. H. Santamore}
\affiliation{ITAMP, Harvard-Smithsonian Center for Astrophysics,
Cambridge, MA 02138}

\affiliation{Department of Physics, Harvard University, Cambridge,
MA 02138}

\affiliation{Department of Physics, California Institute of
Technology, Pasadena, CA 91125}

\author{A. C. Doherty}
\affiliation{Department of Physics, University of Queensland, St.
Lucia, QLD 4072 Australia}

\affiliation{Institute for Quantum Information, California
Institute of Technology, Pasadena, CA 91125}

\affiliation{Department of Physics, California Institute of
Technology, Pasadena, CA 91125}

\author{M. C. Cross}
\affiliation{Department of Physics, California Institute of
Technology, Pasadena, CA 91125}


\begin{abstract}
We investigate a scheme that makes a quantum non-demolition (QND)measurement
of the excitation level of a mesoscopic mechanical oscillator by utilizing the
anharmonic coupling between two beam bending modes. The non-linear coupling
between the two modes shifts the resonant frequency of the readout oscillator
in proportion to the excitation level of the system oscillator. This frequency
shift may be detected as a phase shift of the readout oscillation when driven
on resonance. We derive an equation for the reduced density matrix of the
system oscillator, and use this to study the conditions under which discrete
jumps in the excitation level occur. The appearance of jumps in the actual
quantity measured is also studied using the method of quantum trajectories. We
consider the feasibility of the scheme for experimentally accessible parameters.

\end{abstract}
\date{\today  }
\maketitle

\section{Introduction}

Quantum mechanics tells us that the energy of an oscillator is quantized.
However, an observation of quantum limited mechanical motion in macroscopic
objects has not been possible because the energy associated with individual
phonons is typically much smaller than the thermal energy \cite{C03, R01}.

Advances in nanotechnology have enabled experimenters to build ever smaller
mechanical oscillators with high resonance frequencies and quality factors
\cite{HZMR03}. As an individual phonon energy becomes comparable to or greater
than $k_{\mathrm{B}}T$, quantum effects begin to appear and it should be
possible to realize various quantum phenomena.

In this paper, we investigate the possibility of observing transitions amongst
the Fock states of a mesoscopic mechanical oscillator. To do this requires the
coupling of the system oscillator to a measurement device that sensitively
detects the phonon number of the system oscillator but does not itself change
the excitation level of the oscillator. In the quantum regime it becomes very
important to model the precise way that a quantum system interacts with any
measuring apparatus, as well as with the environment. Specifically, it is
necessary to take into account the measurement backaction and to design the
system-readout interaction so as to allow the best possible measurement of the
desired observable. We will show that it is possible in principle to take
advantage of the non-linear interaction between modes of oscillation of an
elastic beam or beams to track the state of the oscillator as it jumps between
number states due to its coupling to the surrounding thermal environment.

The laws of quantum mechanics tell us that, even in the absence of
instrumental or thermal noise, a measurement will tend to disturb the state of
the measured system. The interaction between the system and the measurement
apparatus means that while information about the measured observable may be
read out from the state of the meter after interacting with the system, the
quantum mechanical uncertainty in the initial state of the meter leads to
random changes in the conjugate observable of the system. This backaction
noise on the conjugate observable is an inevitable result of the very
interaction that allows the measurement to take place. It has long been
recognized that such backaction noise places a fundamental limit on the
sensitivity of physical measurements \cite{BK}. However, the class of
measurements known as quantum non-demolition (QND) measurements partially
circumvents this problem by guaranteeing that the backaction noise does not
affect the results of future measurements of the same quantity. The idea of a
QND measurement is widely discussed in the literature (for example, see
\cite{BK}, \cite{CTDSZ80}, and \cite{WM}). In a QND measurement the
interaction Hamiltonian between system and meter commutes with the internal
Hamiltonian of the system: an ideal QND measurement is \emph{repeatable} since
the backaction noise does not affect the dynamics of the measured observable.
In this paper, we are interested in a QND measurement of phonon number. The
conjugate observable of number is phase, thus, the measurement backaction in
our case will result in diffusion of the phase of the mechanical oscillations.
However, the scheme allows the complete determination of the oscillator
excitation level and thus projects the system onto a number state in an
idealized limit.

The scheme for the QND measurement of phonon number that we consider uses two
anharmonically coupled modes of oscillation of a mesoscopic elastic structure.
The resonant frequencies of these two modes are different. The higher
frequency mode is the system to be measured, while the lower frequency
oscillator serves as the meter (we refer this oscillator as ancilla). The key
idea of the scheme is that, from the point of view of the readout oscillator,
the interaction with the system constitutes a shift in resonance frequency
that is proportional to the time-averaged excitation of the system oscillator.
This frequency shift may be detected as a change in the phase of the ancilla
oscillations when driven on resonance. We show that this scheme realizes an
ideal QND measurement of phonon number in an appropriate limit. To measure the
phase of the ancilla oscillator we imagine a magnetomotive detection scheme so
that the actual physical quantity measured is an electric current that couples
to the ancilla displacement. Thus our task is to understand how the strong
measurement of the current, represented by the von Neumann projection scheme
on the current operator, yields information on the system phonon number, and
in turn affects the dynamics of the system via the indirect coupling through
the ancilla oscillator, and in the presence of the inevitable coupling of the
ancilla and system oscillators to the environment. This QND measurement scheme
where a nonlinear potential provides a phase shift to one oscillator that
reflects the excitation of the other is analogous to the experiment of Peil
and Gabrielse \cite{PG99}, which demonstrated a QND measurement of the
excitation of a single trapped electron. Theoretical discussions of such
approximate QND measurements of the excitation of an oscillator date back at
least to Unruh \cite{U78}.

In Sec.\ \ref{Sec_modelHamiltonian} we introduce our model and construct the
Hamiltonian describing the two oscillators, the magnetomotive coupling, and
the coupling to the environment represented by baths of harmonic oscillators.
For the ancilla oscillator displacement to directly indicate the system phonon
number the time scale of the ancilla dynamics must be much shorter than that
of the system. This actually allows us to adiabatically eliminate the ancilla
operators to obtain dynamical equations for the system alone. Thus, in
Sec.\ \ref{Sec_dynamicaleqn}, we obtain a reduced master equation for the
density matrix of the system oscillator, which allows us to focus on the
physics of the system dynamics. This adiabatic elimination of ancilla degrees
of freedom is often considered in quantum optics. However, the adiabatic
elimination used in quantum optics is at temperature zero, and we need to
reformulate the method for finite temperatures.

Once we know the system dynamics, we next focus on obtaining the experimental
outcome. Quantum mechanics allows us to determine the state of the system
\emph{conditioned} on the measured current $I(t)$. The von Neumann projection
postulate says that after a measurement a quantum system in some possibly
mixed initial state is projected onto the eigenstate corresponding to the
measurement outcome. The continual measurement and projection of the current
$I(t)$ provides accumulating information on the system phonon number, and
correspondingly a projection onto phonon number states. The theory of quantum
trajectories \cite{B86,B87,C93,Wth} has been developed to deal with such
continuous measurements. In Secs.\ \ref{Sec_trajectories} we discuss such
quantum trajectory equations for our system. The method leads to a stochastic
master equation for the system density matrix, where the stochastic component
comes from the particular value of the measured current at each time, which
itself is a stochastic variable since it is the outcome of a quantum
measurement. These equations of motion for the system conditioned on a
particular sequence of measurement results allow us to investigate the
possibility of tracking the evolution of the system as it jumps between number
states due to its interaction with the thermal bath. Some details of the
formulation of the operators describing the measurement current that are
needed to derive the stochastic master equation are given in Appendix
\ref{Subsec:bathfield}.

The main discussion of the physical implications of the model is in
Sec.\ \ref{SubSec_competition} where we consider the parameters that are
necessary to observe the oscillator jump between number states. A reader not
interested in the details of the derivation could read
Sec.\ \ref{SubSec_competition} and following sections after the description of
the model in Sec.\ \ref{Sec_modelHamiltonian}. Finally, in
Sec.\ \ref{SubSec_parameter} we conclude with a discussion of the feasibility
of the scheme based on current technology and future enhancements.

\section{Constructing the Hamiltonian\label{Sec_modelHamiltonian}}

\subsection{The model\label{SubSec:model}}

In this section we introduce the model system and show how the coupling
between the system and ancilla approximates a QND coupling in an appropriate
limit. We then derive equations of motion that take into account the couplings
to the environment and the interactions that drive and monitor the
oscillations of the ancilla.

\begin{figure}[ptbh]
%\begin{figure}[tbp]
\par
\begin{center}
\includegraphics[width=4in]{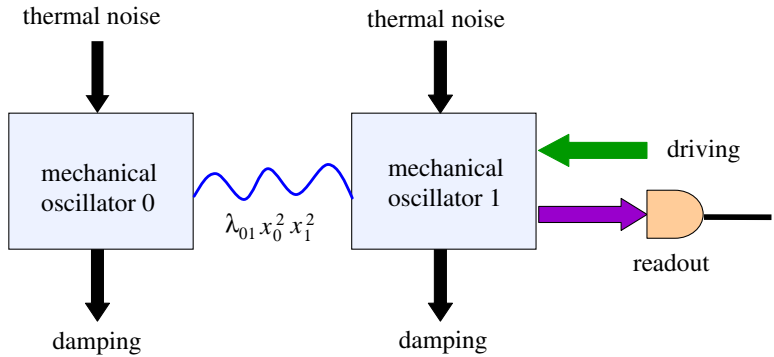}
\end{center}
\caption{{Schematic of a QND measurement using two coupled
mechanical
oscillators.}}%
\label{fig:QNDmodel}%
\end{figure}

Consider a mesoscopic beam with rectangular cross section. There are two
orthogonal flexing modes that are not coupled in the linear elasticity theory,
but are coupled anharmonically. This coupling exists in nature between the two
orthogonal flexing modes of single mechanical beam. However, the coupling can
also be controlled and engineered: a\ similar coupling of bending modes in two
elastic beams has been proposed by Yurke \cite{SP:Y}. In this scheme two
mesoscopic elastic beams with rectangular cross section are connected by a
series of mechanical coupling devices. These devices have the effect of
allowing only one type of strain (the longitudinal stretch) to pass to the
other beam. In this paper we focus on the extent to which such mechanical
devices are able to realize a QND measurement and the constraints this places
on the specifications of the device, and the temperature at which the
experiment is performed.

For convenience, we refer to the system of interest as oscillator $0$ and the
ancilla as oscillator $1$, and the corresponding resonant frequencies of the
two modes as $\omega_{0}$ and $\omega_{1}$, respectively.\ The ancilla is
driven at its resonant frequency, and a measurement apparatus is attached to
the ancilla. The whole structure is kept at a low temperature $T$ such that
$\hbar\omega_{0}\sim k_{B}T$, where $\hbar=h/2\pi$ is Plank's constant, and
$k_{B}$ is Boltzmann's constant. The oscillators are weakly coupled to the
environment. Figure \ref{fig:QNDmodel} shows a schematic of our model.

\subsection{System Hamiltonian\label{SubSec_Hamiltonian}}

Converting the schematic model in the last section into a tractable
mathematical model and obtaining the system dynamics requires some assumptions
and simplifications.

Firstly, we focus on the anharmonic coupling and the limit in which it
satisfies the QND condition. In linear elasticity theory, the two flexing
modes, which are perpendicular to each other, propagate independently without
interacting. Beyond the linear approximation these modes are coupled.
Expansion of the elastic energy with respect to the strain tensor is taken up
to second order in the harmonic approximation. By symmetry the coupling
between the modes first occurs at fourth order, proportional to $x_{0}%
^{2}x_{1}^{2}$. So we expand the anharmonic terms up to quartic order to give%
\begin{equation}
H_{\mathrm{anh}}=\hbar\left(  \tilde{\lambda}_{00}x_{0}^{4}+\tilde{\lambda
}_{11}x_{1}^{4}\right)  +\hbar\tilde{\lambda}_{01}x_{0}^{2}x_{1}^{2},
\label{eq:anharmonicfull}%
\end{equation}
where $\tilde{\lambda}_{ij}$ give the strengths of the nonlinear terms. The
first two terms in Eq.\ (\ref{eq:anharmonicfull}) are internal anharmonic
terms. Under the rotating wave approximation (see below), these terms cause a
shift in the oscillator resonant frequencies and a non-linear phase shift that
depends on intensity (a Kerr non-linearity) resulting in rotational shear of
the state in the phase space of the two oscillators. For the system
oscillator, the small shifts in the energy level spacings are not important,
and can be ignored. The ancilla oscillator is externally driven, and so the
nonlinearity of this oscillator may become large, for example leading to
multistability for large enough drive strengths. We will assume that the drive
strength is kept smaller than this range, so that again the $x_{1}^{4}$
nonlinearity does not play an essential role. However the $x_{0}^{2}x_{1}^{2}$
coupling plays an essential role in coupling the system and ancilla.
Therefore, in the interests of a straightforward presentation we retain the
non-linear coupling given by $\lambda_{01}$ and disregard the nonlinearities
of the system and ancilla internal Hamiltonians. A detailed analysis including
non-interacting nonlinearities and detuning in Ref.\ (\cite{qph04}) has shown
that in the regime of strong damping of the ancilla that we will mostly
consider the effect of these anharmonic terms will be negligible for small detuning.

In terms of creation and annihilation operators, the Hamiltonian is now
$H=H_{0}+V_{\mathrm{anh}}$, with $H_{0}$ the harmonic part%
\begin{equation}
H_{0}=\hbar\omega_{0}a_{0}^{\dagger}a_{0}+\hbar\omega_{1}a_{1}^{\dagger}a_{1},
\label{eq:freeH}%
\end{equation}
and $V_{\mathrm{anh}}$ the anharmonic coupling%
\begin{equation}
V_{\mathrm{anh}}=\frac{1}{4}\hbar\lambda_{01}\left(  a_{0}^{\dagger}%
+a_{0}\right)  ^{2}\left(  a_{1}^{\dagger}+a_{1}\right)  ^{2}. \label{eq:intH}%
\end{equation}
We have defined the standard raising and lowering operators for the
oscillators
\begin{equation}
a_{i}=\sqrt{m_{i}\omega_{i}/2\hbar}x_{i}+i\sqrt{1/2\hbar m_{i}\omega_{i}}%
p_{i},
\end{equation}
and $a_{i}^{\dagger}$ is the Hermitian conjugate of $a_{i}$. So far we have
ignored any coupling of the two oscillators to the environment so as to focus
on the interaction of the two oscillators.

In order to perform a QND measurement of $a_{0}^{\dagger}a_{0}$ the
Hamiltonian of the oscillators $H$ should satisfy the QND condition%
\begin{equation}
\lbrack a_{0}^{\dagger}a_{0},H_{0}+V_{\mathrm{anh}}]=0. \label{QNDcondition}%
\end{equation}
To show that this condition is satisfied in an appropriate limit it is useful
to move into an interaction picture with respect to $H_{0}$. If the
frequencies of the two oscillators satisfy $\omega_{0}-\omega_{1}\gg
\lambda_{01}$ and $\omega_{i}\gg\lambda_{01}$, then the time-dependent terms
in $V_{\mathrm{anh}}(t)$ lead to rapid, small amplitude oscillations of
$a_{i}$ that essentially average to zero over the time scales for which the
non-linearity $\lambda_{01}$ is relevant. If we admit a time coarse-graining
over times longer than the mechanical oscillation period we may ignore the
rapidly oscillating terms, an approximation known as the rotating wave
approximation (RWA). Another intuitive explanation for the rotating wave
approximation is that the condition $\omega_{0}-\omega_{1}\gg\lambda_{01}$
means that the differences in energy are so large that the energy
non-conserving transitions are strongly suppressed.

Disregarding the energy non-conserving terms in the Hamiltonian and then
absorbing constant corrections to the system and ancilla oscillation frequency
into the definition of $\omega_{0}$ and $\omega_{1}$, we obtain
\begin{equation}
V_{\mathrm{anh}}^{\mathrm{RWA}}(t)=\hbar\lambda_{01}a_{0}^{\dagger}a_{0}%
a_{1}^{\dagger}a_{1}.
\end{equation}
The constant term has been disregarded since it merely provides an overall
phase. Note that having made the rotating wave approximation the anharmonic
coupling term commutes with the observable $a_{0}^{\dagger}a_{0}$, and so a
QND measurement can be achieved under the condition $\omega_{0}-\omega_{1}%
\gg\lambda_{01}$ \footnote{Actually, it is not necessary to make the rotating
wave approximation with respect to both oscillators as the interaction
$V_{\mathrm{anh}}$ becomes QND after the rotating wave approximation made in
an interaction picture with respect to $\hbar\omega_{0}a_{0}^{\dagger}a_{0}$
only. However, we make the further rotating wave approximation to simplify the
later development.}. Returning to the Schr\"{o}dinger picture the Hamiltonian
$H$ now can be written as%
\begin{equation}
H^{\mathrm{RWA}}=\hbar\omega_{0}a_{0}^{\dagger}a_{0}+\hbar\omega_{1}%
a_{1}^{\dagger}a_{1}+\hbar\lambda_{01}a_{0}^{\dagger}a_{0}a_{1}^{\dagger}%
a_{1}. \label{eq:HRWA}%
\end{equation}

In the above rotating wave Hamiltonian an excitation of the system oscillator
leads to a frequency shift of the ancilla oscillator. To detect the system
excitation level, the ancilla is driven on resonance and the phase shift of
the oscillations is measured. The driving of the ancilla may be written in
terms of term in the Hamiltonian in the Schr\"{o}dinger picture
\begin{equation}
H_{\text{\textrm{drive}}}=2\hbar E\left(  a_{1}+a_{1}^{\dagger}\right)
\cos\omega_{1}t, \label{eq:drivecurrentH}%
\end{equation}
where the parameter $E$ is used to characterize the strength of the drive. In
the interaction picture using the RWA for $\omega_{1}>E$, we get
\begin{equation}
H_{\text{\textrm{drive}}}^{\mathrm{RWA}}=\hbar E\left(  a_{1}+a_{1}^{\dagger
}\right)  . \label{eq:Hdrive}%
\end{equation}

Now we add the coupling of thermal baths to the system and ancilla. We employ
a standard technique and model the thermal baths (the surrounding environment)
as an infinite number of harmonic oscillators. The thermal baths are linearly
coupled to the system or ancilla by coordinate-coordinate coupling,
\textit{i.e.}, $\sum_{j}A_{ij}x_{i}x_{j}$ where $x_{i}$ is the system or
ancilla coordinate, $x_{j}$\ is the coordinate of an oscillator in the bath,
with the index $j$ corresponding to different bath oscillators. We will again
use the rotating wave approximation for the coupling since the couplings are weak.

The nature of the coupling with the measurement instrument depends on the
measurement scheme. Here we adopt a magnetomotive detection scheme suggested
by Yurke \textit{et al.}\ \cite{YGPB94, CR96, CR99}. A metallized conducting
surface on the ancilla oscillator develops an electromotive force across it
due to a perpendicular magnetic field and the oscillation of the beam. The
voltage developed depends on
\begin{equation}
V=lB\frac{dx_{1}}{dt}, \label{eq:signal}%
\end{equation}
where $V$ is the voltage, $B$ is the magnetic field, the conductor is of
length $l$ and $x_{1}$ is the displacement of the beam from its equilibrium
position. Depending on the resistance $R$ of the conducting strip and
remainder of the circuit this will result in a current that is then measured.

In order to quantize this measuring device we follow the standard practice in
quantum electronics and model this resistance by a semi-infinite transmission
line\cite{L}. This model has been used in the context of mechanical
measurements (see \cite{YD,PM01}) and is in fact mathematically the same as
the 'Rubin model' \cite{R63,W}. This is certainly a simplified model of an
actual detection circuit which essentially assumes that the noise in the
circuit is broad-band and Gaussian. The transmission line will be considered
to be at a temperature corresponding to the effective noise temperature of the
detection circuit and this noise will affect both the sensitivity and the
heating of the ancilla and system oscillators. More realistic quantum
mechanical models of amplifier circuits can be found in \cite{C98} for
example. Our final model for the QND set-up will be fairly robust to the
precise detection circuit. The resulting current operator is
\begin{equation}
I\propto\sum_{n}b_{\mathrm{d},n}+b_{\mathrm{d},n}^{\dagger}%
,\label{eq:bathcurrent}%
\end{equation}
where $b_{\mathrm{d},n}$ are the lowering operators for the modes of the
transmission line and the proportionality constant, which is not important for
our results, depends on the circuit parameters. For a linearly coupled
system-bath measurement within the rotating wave approximation, the
Hamiltonian describing the coupling between each measurement current mode and
the ancilla is proportional to $b_{\mathrm{d},n}^{\dagger}a_{1}+b_{\mathrm{d}%
,n}a_{1}^{\dagger}$. The coupling to the thermal bath modes has the same
mathematical structure. In the rotating wave approximation, the difference
between a coordinate-coordinate coupling, and a momentum-coordinate coupling
can be absorbed into the definition of the phase of the various raising and
lowering operators. As is usually done, for later convenience we will include
a phase factor of $\pi/2$ so that the coupling to the baths and measurement
current takes the latter form.

The final Hamiltonian for our model is then%
\begin{align}
H  &  =\hbar\omega_{0}a_{0}^{\dagger}a_{0}+\hbar E\left(  a_{1}+a_{1}%
^{\dagger}\right)  +\hbar\sum_{\mathrm{s}}\sum_{n}^{\infty}\omega
_{\mathrm{s},n}b_{\mathrm{s},n}^{\dagger}b_{\mathrm{s},n}\nonumber\\
&  +\hbar\lambda_{01}a_{0}^{\dagger}a_{0}a_{1}^{\dagger}a_{1}+i\hbar\left(
\Theta^{\dagger}a_{0}-\Theta a_{0}^{\dagger}\right)  +i\hbar\left(
\Gamma^{\dagger}a_{1}-\Gamma a_{1}^{\dagger}\right)  +i\hbar\left(
D^{\dagger}a_{1}-D\left(  t\right)  a_{1}^{\dagger}\right)  ,
\label{eq:fullHamiltonian}%
\end{align}
where $\Gamma,D,\Theta$ have the form, $\sum_{\mathrm{n}}^{\infty
}g_{\mathrm{s}}\left(  \omega_{n}\right)  b_{\mathrm{s},n},$ and the index
$\mathrm{s}$ denotes the three different baths: the thermal bath coupled to
the system (0), the thermal bath coupled to the ancilla (1), and the
measurement bath coupled to the ancilla (d). The strength of the coupling to
the bath modes is given by the coefficients $g_{\mathrm{s}}\left(  \omega
_{n}\right)  $. Later, we will derive the relationship between these
coefficients and the corresponding oscillator damping rates or quality factors.

\section{Dynamics of the system\label{Sec_dynamicaleqn}}

We use the dynamics described by Eq.\ (\ref{eq:fullHamiltonian}) to understand
the measurement process. In this section we first find a master equation that
describes the evolution of the system and ancilla alone without explicitly
describing the state of the environment. Secondly we further simplify this
equation by making use of the difference in time-scales between the system and
ancilla to obtain a master equation for the system oscillator alone by means
of adiabatic elimination. This allows us to study the effect of the QND
measurement coupling on the system.

\subsection{Master equation\label{SubSec_mastereqn}}

We develop a master equation for the density operator of the system alone by
integrating out the bath degrees of freedom. Because we are interested in
high-Q oscillators weakly coupled to the baths we employ a rotating wave
approximation and the Markov approximation that the memory time scale of the
bath is short. In this regime the rotating wave master equation accurately
describes the dynamics on time-scales longer than an oscillation period, and
the resulting master equation preserves the positivity of the density matrix.
The derivation of such master equations is widely discussed in the literature,
see for example Carmichael \cite{C}, Walls and Milburn \cite{WM} and Caldeira
and Leggett \cite{CL83}. Note that Caldeira and Leggett do not make the
rotating wave approximation (which we adopt from the quantum optics
literature) but instead make a high-temperature approximation; the two
equations agree in the overlap of their domain of validity (high temperature
and weak coupling to the bath). However, the Caldeira-Leggett master equation
can only be guaranteed to preserve the positivity of the density operator in
the limit of high temperature.

Assuming that the environment and measurement baths are in thermal
equilibrium, the master equation for the reduced density operator $\rho$
describing the state of the system and ancilla in the interaction picture
takes what is known as Lindblad form
\begin{align}
\frac{d\rho}{dt}  &  =-\frac{i}{\hbar}\left[  \hbar E\left(  a_{1}^{\dagger
}+a_{1}\right)  +\hbar\lambda_{01}a_{0}^{\dagger}a_{0}a_{1}^{\dagger}%
a_{1},\rho\right] \nonumber\\
&  +\nu(N_{0}+1)\mathcal{D}\left[  a_{0}\right]  \rho+\nu N_{0}\mathcal{D}%
\left[  a_{0}^{\dagger}\right]  \rho+\kappa(N_{1}+1)\mathcal{D}\left[
a_{1}\right]  \rho+\kappa N_{1}\mathcal{D}\left[  a_{1}^{\dagger}\right]
\rho, \label{eq:master eqn}%
\end{align}
where
\[
\mathcal{D}\left[  x\right]  \rho=2x\rho x^{\dagger}-x^{\dagger}x\rho-\rho
x^{\dagger}x,
\]
and $N_{\mathrm{i}}$ are the Bose-Einstein factors at frequency $\omega_{0}$
for $N_{0}$ and $\omega_{1}$ for $N_{1}$. The first term in
Eq.\ (\ref{eq:master eqn}) involving the commutator describes the coherent
driving of the ancilla oscillator and the non-linear coupling between the two
oscillators in the rotating wave approximation. The remaining terms describe
the dissipative interactions with the various baths. The coefficient $\nu$ is
\[
\nu\equiv\pi\varrho_{\mathrm{b0}}\left(  \omega_{0}\right)  \left\vert
g_{\mathrm{b0}}\left(  \omega_{0}\right)  \right\vert ^{2},
\]
where $\varrho_{\mathrm{b0}}\left(  \omega_{0}\right)  $ is the density of
states of the bath coupled to the system at frequency $\omega_{0}$. It can be
experimentally obtained from the quality factor of the system oscillator
$Q_{0}$ as $\nu=\omega_{\mathrm{0}}/2Q_{\mathrm{0}}$. The coefficient $\kappa$
is the corresponding damping rate of the ancilla, with contributions $\eta$
from the coupling to the environment and $\mu$ from the measurement bath. Both
these rates can be expressed in terms of the bath density of states at
frequency $\omega_{1}$ in exactly the same way as for $\nu$, and $\kappa
=\eta+\mu$. The terms containing a factor $(N_{i}+1)$ describe the spontaneous
and stimulated emission of phonons into the thermal bath while the ones with
$N_{i}$ correspond to absorption of phonons.

The master equation Eq.\ (\ref{eq:master eqn}) can in principle be numerically
integrated. However we will make some further approximations in order to
derive a master equation for the system dynamics alone and show that in some
limit the readout system coupling results in the phase diffusion that is
required as the backaction for the QND measurement, with no extra noise above
this quantum limit. To do this we assume that the ancilla is strongly damped.
In this limit the ancilla relaxes rapidly to a state consistent with the
instantaneous system state. As a result its dynamics are slaved to the system
oscillator and can in fact be eliminated from the equations of motion.
Experimentally this is the limit in which the displacement of the ancilla
directly reflects the system behavior. This \emph{adiabatic elimination} is
described in the following subsection. The final result of this analysis is
Eq.\ (\ref{eq:redmastereqn}) below.

Note that $\nu,\kappa$ are the widths of the oscillator resonances, and these
should be taken into account when assessing the validity of the rotating wave
approximation. In the presence of the coupling to the baths, the rotating wave
approximation is only valid if $\omega_{0}-\omega_{1}$ is much greater than
the line width of the oscillators, \textit{i.e.}, $\omega_{0}-\omega_{1}\gg
\nu,\kappa$. This condition can be understood as not allowing
non-energy-conserving transfers of a phonon between the two oscillators.

\subsection{Adiabatic elimination\label{SubSec_adiabatic}}

For a strongly damped ancilla ($\kappa\gg\nu$) the driven ancilla rapidly
relaxes to a state that oscillates with a phase determined by the current
system phonon number. In the interaction picture this is a displaced thermal
state: a state with variance of position and momentum equal to those of a
thermal state but with non-zero expectation values of position and momentum
consistent with the driving and damping of the oscillator. It will be useful
to transform the equations of motion in such a way as to make a perturbative
expansion around this steady state. The basic idea is to transform the origin
of phase space such that the ancilla steady state for the transformed master
equation is a thermal state. This transformation will essentially remove the
driving term in the master equation. This is the approach of Wiseman and
Milburn \cite{WM00} who study adiabatic elimination in a similar model. While
they assume zero temperature and therefore end up with a perturbation
expansion about the displaced ancilla ground state we must generalize their
techniques to finite temperature.

Following Wiseman and Milburn, we use the displacement operator,
$D(\alpha)=\exp\left[  \alpha a_{1}^{\dagger}-\alpha^{\ast}a_{1}\right]  $
with $\alpha=-iE/\kappa$. The transformed system state is $\tilde{\rho}\equiv
D(\alpha)\rho D(\alpha)^{\dagger}$ and we may write the master equation for
$\tilde{\rho}$ as%
\begin{align}
\dot{\tilde{\rho}}  &  =D(\alpha)\dot{\rho}D(\alpha)^{\dagger}\nonumber\\
&  =-i\left\vert \alpha\right\vert ^{2}\lambda_{01}\left[  a_{0}^{\dagger
}a_{0},\tilde{\rho}\right] \nonumber\\
&  -i\lambda_{01}\left[  a_{0}^{\dagger}a_{0}a_{1}^{\dagger}a_{1},\tilde{\rho
}\right]  -i\lambda_{01}\left[  a_{0}^{\dagger}a_{0}\left(  \alpha
a_{1}^{\dagger}+\alpha^{\ast}a_{1}\right)  ,\tilde{\rho}\right] \nonumber\\
&  +\kappa(N_{1}+1)\left(  2a_{1}\tilde{\rho}a_{1}^{\dagger}-a_{1}^{\dagger
}a_{1}\tilde{\rho}-\tilde{\rho}a_{1}^{\dagger}a_{1}\right)  +\kappa
N_{1}\left(  2a_{1}^{\dagger}\tilde{\rho}a_{1}-a_{1}a_{1}^{\dagger}\tilde
{\rho}-\tilde{\rho}a_{1}a_{1}^{\dagger}\right) \nonumber\\
&  +\nu(N_{0}+1)\left(  2a_{0}\tilde{\rho}a_{0}^{\dagger}-a_{0}^{\dagger}%
a_{0}\tilde{\rho}-\tilde{\rho}a_{0}^{\dagger}a_{0}\right)  +\nu N_{0}\left(
2a_{0}^{\dagger}\tilde{\rho}a_{0}-\tilde{\rho}a_{0}a_{0}^{\dagger}-a_{0}%
a_{0}^{\dagger}\tilde{\rho}\right)  . \label{eq:D-master-eqn}%
\end{align}
In this master equation the excitation of the ancilla oscillations leads to a
frequency shift of the system oscillator described by the first three terms on
the right hand side of this equation. The first term is due to the classical
mean value of the ancilla oscillator energy and is just a constant shift in
the system oscillation frequency. We may move to an interaction picture at
this shifted frequency, the most convenient interaction picture in which to
perform the adiabatic elimination. The next two terms describe the effect of
the fluctuations in the ancilla excitation. The thermal bath coupling terms
(the last four groups of terms) are the same as before.

The adiabatic elimination will hold when the terms proportional to $\kappa$ in
Eq.\ (\ref{eq:D-master-eqn}) dominate in the ancilla dynamics. Thus, the
adiabatic elimination is valid in a strongly damped regime such that
\begin{equation}
\frac{\lambda_{01}\left\vert \alpha\right\vert }{\kappa},\;\frac{\nu}{\kappa
}\;\simeq\epsilon\ll1. \label{eq:constrains}%
\end{equation}
We are assuming that the ancilla oscillator relaxes faster than the system
oscillator as well as that the non-linear dynamics are weak compared to the
damping of the ancilla oscillator. For the consistency of the following
treatment it will also be necessary to have $\lambda_{01}N_{1}/\kappa
\simeq\epsilon^{2}$. This requirement follows from the second term (the
non-linear coupling term) on the right hand side of the master equation. This
constraint can be achieved consistent with Eq.\ (\ref{eq:constrains}), for
example, by leaving $\alpha$ finite and choosing $N_{1},\lambda_{01}%
/\kappa\simeq\epsilon$, a regime of low temperature and moderate
non-linearity. The approximations are also valid at arbitrary temperature in
the limit of strong driving and weak non-linearity such that $\lambda
_{01}/\kappa\simeq\epsilon^{2}$ and $\alpha\simeq\epsilon^{-1}$ hold
\footnote{However, the driving strength is chosen in such a way to avoid the
bi-stability region caused by the anharmonicity of the oscillator, see
\ref{SubSec_parameter}.}. Here the scaling of the driving strength is chosen
to preserve the measurement sensitivity which will scale with $\lambda
_{01}\alpha/\kappa$. In this regime the frequency shift of the system
oscillator becomes large.

As mentioned above, in this displaced frame the state of the readout
oscillator is close to a thermal state and we expand $\tilde{\rho}$ in the
form
\begin{align}
\tilde{\rho}  &  =\rho_{0}\otimes\rho_{N_{1}}+\rho_{1}\otimes a_{1}^{\dagger
}\rho_{N_{1}}+\rho_{1}^{\dagger}\otimes\rho_{N_{1}}a_{1}+\rho_{2}\otimes
a_{1}^{\dagger}\rho_{N_{1}}a_{1}\nonumber\\
&  +\rho_{2^{\prime}}\otimes a_{1}^{\dagger2}\rho_{N_{1}}+\rho_{2^{\prime}%
}^{\dagger}\otimes\rho_{N_{1}}a_{1}^{2}+O\left(  \epsilon^{3}\right)  .
\label{eq:tot-densityl}%
\end{align}
Here the $\rho_{i},i=0,1,2\ldots$\ act on the system oscillator and the
subscripts indicate orders of magnitude in $\epsilon$. The scalings of the
different parameters with $\epsilon$ have been chosen to guarantee the
consistency of the expansion. The quantity $\rho_{N_{1}}$ is the thermal
density matrix for the ancilla, which in terms of the average excitation
number $N_{1}$\ is
\begin{equation}
\rho_{N_{1}}=\sum_{n=0}\frac{1}{N_{1}+1}\left(  \frac{N_{1}}{N_{1}+1}\right)
^{n}|n\rangle\langle n|. \label{eq:thermal state}%
\end{equation}
This is the steady state of the master equation for an oscillator coupled to a
thermal bath with temperature given by $N_{1}$ We have restricted
Eq.\ (\ref{eq:tot-densityl}) to normal ordered terms using the following
identities which can be proved from this expression for $\rho_{N_{1}}$:%
\begin{align}
\rho_{N_{1}}a_{1}^{\dagger}  &  =\frac{N_{1}}{N_{1}+1}a_{1}^{\dagger}%
\rho_{N_{1}},\\
a_{1}\rho_{N_{1}}  &  =\frac{N_{1}}{N_{1}+1}\rho_{N_{1}}a_{1}. \label{eq:ids3}%
\end{align}
These normal ordering identities are the key to generalizing the arguments of
Wiseman and Milburn to the finite temperature case. Using $\mathrm{Tr}%
_{1}(\rho_{N_{1}})=1$ and $\mathrm{Tr}_{1}(a_{1}^{\dagger}\rho_{N_{1}}%
a_{1})=N_{1}+1$, it can be seen that the system density matrix after tracing
out the ancilla state is
\begin{equation}
\rho_{\mathrm{s}}=\mathrm{Tr}_{1}\left\{  \tilde{\rho}\right\}  =\rho
_{0}+(N_{1}+1)\rho_{2}. \label{eq:Tr rho}%
\end{equation}

Now we substitute Eq.\ (\ref{eq:tot-densityl}) into
Eq.\ (\ref{eq:D-master-eqn}) and, using the normal ordering identities, derive
equations of motion for the operators $\rho_{i}$, retaining terms in the
evolution of $\rho_{0}$ and $\rho_{2}$ up to second order in $\epsilon$:
\begin{align}
\dot{\rho}_{0}=  &  -i\lambda_{01}\left[  \alpha^{\ast}a_{0}^{\dagger}%
a_{0}\rho_{1}-\alpha\rho_{1}^{\dagger}a_{0}^{\dagger}a_{0}\right]
+2\kappa(N_{1}+1)\rho_{2}+\mathcal{L}_{0}\rho_{0}+\kappa\mathcal{O}\left(
\epsilon^{3}\right)  ,\label{eq:rho_0/dt}\\
\dot{\rho}_{1}=  &  -i\lambda_{01}a_{0}^{\dagger}a_{0}\alpha\rho_{0}%
+i\lambda_{01}\alpha\frac{N_{1}}{N_{1}+1}\rho_{0}a_{0}^{\dagger}a_{0}%
-\kappa\rho_{1}+\kappa\mathcal{O}\left(  \epsilon^{2}\right)  ,\\
\dot{\rho}_{2}=  &  -i\lambda_{01}a_{0}^{\dagger}a_{0}\left[  \alpha\rho
_{1}^{\dagger}+\alpha^{\ast}\frac{N_{1}}{N_{1}+1}\rho_{1}\right]
+i\lambda_{01}\left[  \alpha\frac{N_{1}}{N_{1}+1}\rho_{1}^{\dagger}%
+\alpha^{\ast}\rho_{1}\right]  a_{0}^{\dagger}a_{0}\nonumber\\
&  -2\kappa\rho_{2}-i\lambda_{01}\frac{N_{1}}{N_{1}+1}\left[  a_{0}^{\dagger
}a_{0},\rho_{0}\right]  +\kappa\mathcal{O}\left(  \epsilon^{3}\right)  ,
\label{eq:rho_2/dt}%
\end{align}
where $\mathcal{L}_{0}\rho_{0}$ refers to the damping of the system oscillator
described by the last two terms in Eq.~(\ref{eq:D-master-eqn}). When $\kappa$
is large, the equation for $\rho_{1}$ is strongly damped and quickly decays to
the steady state. So we perform adiabatic elimination by setting $\dot{\rho
}_{1}=0$ and obtaining an expression for $\rho_{1}$
\begin{equation}
\rho_{1}=-i\frac{\lambda_{01}}{\kappa}\left[  \alpha a_{0}^{\dagger}a_{0}%
\rho_{0}-\alpha\frac{N_{1}}{N_{1}+1}\rho_{0}a_{0}^{\dagger}a_{0}\right]
+\mathcal{O}\left(  \epsilon^{2}\right)  . \label{eq:rho1}%
\end{equation}
Substituting Eq.\ (\ref{eq:rho1}) into Eqs.\ (\ref{eq:rho_0/dt},
\ref{eq:rho_2/dt}) and using the definition of the reduced density matrix
Eq.\ (\ref{eq:Tr rho}) we find, up to second order in $\epsilon$, the master
equation for the reduced density matrix
\begin{align}
\dot{\rho}_{\mathrm{s}}  &  =-\frac{\lambda_{01}^{2}\left\vert \alpha
\right\vert ^{2}(2N_{1}+1)}{\kappa}\left[  a_{0}^{\dagger}a_{0},\left[
a_{0}^{\dagger}a_{0},\rho_{\mathrm{s}}\right]  \right]  -i\left\{  \omega
_{0}+\lambda_{01}(\left\vert \alpha\right\vert ^{2}+N_{1})\right\}  \left[
a_{0}^{\dagger}a_{0},\rho_{\mathrm{s}}\right] \nonumber\\
&  +\nu\left(  N_{0}+1\right)  \left(  2a_{0}\rho_{\mathrm{s}}a_{0}^{\dagger
}-a_{0}^{\dagger}a_{0}\rho_{\mathrm{s}}-\rho_{\mathrm{s}}a_{0}^{\dagger}%
a_{0}\right)  +\nu N_{0}\left(  2a_{0}^{\dagger}\rho_{\mathrm{s}}a_{0}%
-\rho_{\mathrm{s}}a_{0}a_{0}^{\dagger}-a_{0}a_{0}^{\dagger}\rho_{\mathrm{s}%
}\right)  . \label{eq:redmastereqn}%
\end{align}
This is the main result of this section. Note that the effect of the adiabatic
elimination has essentially been to replace $a_{1}$ by $\lambda_{01}%
|\alpha|a_{0}^{\dagger}a_{0}/\kappa$, an indication that by measuring the
ancilla oscillations it will be possible to obtain information about the
system phonon number.

\section{Quantum trajectories\label{Sec_trajectories} \ }

Equation (\ref{eq:redmastereqn}) describes the statistical behavior of the
system due to the coupling to the thermal bath and the indirect coupling to
the ancilla thermal bath and the measurement bath, but does not tell us how
the measured current reflects the system state, or about the correlations of
the system dynamics with particular measurement outcomes. In this section we
derive an equation of motion for the state of the system conditioned on a
particular sequence of measurement outcomes. This equation is termed a
\textit{quantum trajectory}\cite{B86,B87,C93,Wth}, and results from
continually projecting onto eigenstates of the current. Since the current
effectively measures phonon number, this measurement process will tend to
force the system towards a pure number state that is consistent with the
measurement current. The time scale for this to occur will depend on the
coupling of the system to the measurement apparatus, which is in turn
connected to the sensitivity of the measurement. On the other hand, the
coupling of the system to a thermal bath will lead it to absorb and emit
energy from the bath. Thus, in order to determine which number state the
system is in and track its evolution, it must be possible to distinguish
between one number state and the next in a time that is short compared to the
time scale over which phonons are absorbed from and emitted into the thermal bath.

A\ quantum trajectory is constructed as follows. Over each infinitesimal time
interval, the system and the measurement bath states become weakly entangled
via the interaction Hamiltonian. As a consequence, at each time instant, the
state of the system influences the distribution of the possible values of the
current $I$ that may be obtained in the measurement. In turn, von Neumann
projections of the entangled states allow us to calculate the effect of the
measurement of the current on the system state. The appropriate projection is
onto the current eigenvector corresponding to the measured current value. This
results in a stochastic master equation for the state evolution. To implement
the quantum trajectory approach we perform a simulation by picking the
measurements $I(t)$ from the correct probability distribution and following
the corresponding evolution of the system state. The $I(t)$ curve produced by
such a simulation is representative of a single experimental run, and is a
useful predictor of what the experimentalist might see. There will be a signal
contribution that reflects the system state, as well as a white noise
background arising from both thermal and quantum noise.

%The result for the density matrix on the other hand, describes the
%ensemble of all experiments consistent with the initial conditions
%and the particular $I(t)$.

\subsection{Description of the measurement\label{SubSec_stochastic term}}

While quantum trajectories are discussed in general at zero temperature in the
quantum optics literature, Wiseman has discussed the quantum trajectory
equations for homodyne detection at finite temperature (Ref.\ \cite{Wth},
\S \ 4.4.1). The demodulated current that reflects changes in the phase of the
ancilla oscillation in our setup is mathematically analogous to homodyne
detection at finite temperature, and so we can adopt these results here.

The measurement bath is described by the Boson operators $b_{d,n}$. Since the
measurement bath is assumed to be large, the finely spaced modes with a smooth
density of states leads to a short memory time, a result known as the Markov
limit. To exploit this is is useful to introduce a global bath operator which
captures the combination of bath modes that interact with the ancilla
oscillating at frequency $\omega_{1}$ at time $t$ (see Appendix
\ref{Subsec:bathfield} for the derivation of these results)%
\begin{equation}
B_{t}=\frac{1}{\sqrt{2\pi\rho_{\mathrm{d}}(\omega_{1})}g_{\mathrm{d}}\left(
\omega_{1}\right)  }\sum_{n}g_{\mathrm{d}}\left(  \omega_{n}\right)
b_{\mathrm{d},n}e^{-\iota\left(  \omega_{n}-\omega_{1}\right)  t},
\label{eq:Bt}%
\end{equation}
and has time-local commutation rules in the Markov approximation%
\begin{equation}
\left[  B_{t},B_{t^{\prime}}^{\dagger}\right]  =\delta\left(  t-t^{\prime
}\right)  . \label{eq:Btcommutation}%
\end{equation}
The operator $B_{t}$ should be considered to be a linear combination of
Schr\"{o}dinger picture operators, with the phase factors of the coefficients
depending on the parameter $t$~\cite{GCZ92}. In quantum optics this is termed
the input field operator and, roughly speaking, describes the combination of
bath modes interacting with the system at time $t$. In terms of these
operators the current Eq.\ (\ref{eq:bathcurrent}) (appropriately scaled to
remove proportionality constants) is%
\begin{equation}
I(t)=B_{t}+B_{t}^{\dagger},
\end{equation}
and the interaction Hamiltonian between the ancilla and the measurement bath
in the interaction picture is%
\begin{equation}
H_{\mathrm{int}}^{\mathrm{I}}(t)=-i\hbar\sqrt{2\mu}(B_{t}a_{1}^{\dagger}%
-B_{t}^{\dagger}a_{1}). \label{eq:H_I}%
\end{equation}
with $\mu=\pi\varrho_{\mathrm{d}}\left(  \omega_{1}\right)  \left\vert
g_{\mathrm{d}}\left(  \omega\right)  \right\vert ^{2}$ the ancilla damping
rate coming from the measurement bath coupling as before.

The idea of the calculation is to consider the interaction of the ancilla with
the bath at time $t$, represented by the operator $B_{t}$, over a small time
interval $\Delta t$. It is supposed that each \textquotedblleft
element\textquotedblright\ in the time sequence of the bath $B_{t}$ is
initially described by a thermal state. Over the interval $\Delta t$ the
ancilla and bath states become weakly entangled. Measurement of the current
(\textit{i.e.}, the bath operator $B_{t}+B_{t}^{\dagger}$) then finds a value
of the current equal to an eigenvalue $I$ of the current operator, with the
corresponding eigenstate $\left\vert I\right\rangle $, with a probability
distribution $P(I)$ given by the density matrix of the entangled state in the
usual way.%
\begin{equation}
P(I)=\left\langle I\left\vert \rho\left(  t+\Delta t\right)  \right\vert
I\right\rangle .
\end{equation}
The measurement also projects the density matrix onto the eigenstate
$\left\vert I\right\rangle $
\begin{equation}
\rho\rightarrow\frac{\left\vert I\right\rangle \left\langle I\left\vert
\rho\left(  t+\Delta t\right)  \right\vert I\right\rangle \left\langle
I\right\vert }{\left\langle I\left\vert \rho\left(  t+\Delta t\right)
\right\vert I\right\rangle }. \label{eq:timeevolve}%
\end{equation}
Since the value of the current measured is a stochastic variable, this
projection adds a stochastic component to the evolution of the density matrix.

To follow the evolution over a time $\Delta t$ it is useful to introduce the
normalized operator%
\begin{equation}
\Delta B=\left[  \int_{0}^{\Delta t}B_{t}dt\right]  /\sqrt{\Delta t}\simeq
B_{t}\sqrt{\Delta t},
\end{equation}
which satisfies the commutation rule%
\begin{equation}
\left[  \Delta B\left(  t\right)  ,\Delta B^{\dagger}\left(  t\right)
\right]  =1.
\end{equation}
At time $t$ the density matrix representing the ancilla and the segment of the
measurement bath represented by $\Delta B\left(  t\right)  $ can be written as
a direct product of the system plus ancilla $\rho\left(  t\right)  $ and bath
$\rho_{\mathrm{b}}\left(  t\right)  $ density matrices%
\begin{equation}
\bar{\rho}\left(  t\right)  =\rho\left(  t\right)  \otimes\rho_{\mathrm{b}%
}\left(  t\right)  ,
\end{equation}
and $\rho_{\mathrm{b}}\left(  t\right)  $ is a thermal state. To lowest order
the evolution under the interaction Eq.\ (\ref{eq:H_I}) gives%
\begin{equation}
\bar{\rho}\left(  t+dt\right)  =\rho\left(  t\right)  \otimes\rho_{\mathrm{b}%
}\left(  t\right)  +\sqrt{2\mu}\sqrt{\Delta t}\left[  \Delta B^{\dagger}%
a_{1}-a_{1}^{\dagger}\Delta B,\rho\left(  t\right)  \otimes\rho_{\mathrm{b}%
}\left(  t\right)  \right]  +\mathcal{O}\left(  \Delta t\right)  .
\label{eq:rho(t+dt)before}%
\end{equation}
The second term on the right hand side of this equation is the leading order
term in the weak entangling of the state, and will lead, after projection, to
the stochastic part of the density matrix evolution. Using the $\Delta B$
notation has made the $\mathcal{O}\left(  \sqrt{\Delta t}\right)  $ size of
this term explicit. To derive the deterministic part of the evolution equation
we would need to keep the $\mathcal{O}\left(  \Delta t\right)  $ terms, but
since these are already known (the master equation in Lindblad form) we will
not do this here.

The scheme is now to project this density matrix onto an eigenstate of $\Delta
B+\Delta B^{\dagger}$ chosen with a probability given by $\bar{\rho}\left(
t+dt\right)  $. Because of the weak coupling of the bath with the system, this
will give a small additional contribution (actually proportional to
$\sqrt{\Delta t}$) to the system density matrix depending on the value of the
current measured. Since the combination of operators $\Delta B+\Delta
B^{\dagger}$ is just the displacement\ $X$ of the harmonic oscillator
represented by the operator $\Delta B$, this projection is most easily done by
first transforming the state of the bath into a Wigner function representation
(see for example \cite{Ga}). At time $t$, the bath oscillator described by
$\Delta B(t)$ is in a thermal state and the distribution of $X$ is a Gaussian
centered at $X=0$ and with width $2N_{1}+1=\coth\left(  \hbar\omega
_{1}/2k_{\mathrm{B}}T\right)  $. Following the evolution of the state shows
that at time $t+\Delta t$ and to $\mathcal{O}\left(  \sqrt{\Delta t}\right)  $
the distribution of $X$ after the evolution corresponding to the operation
Eqs.\ (\ref{eq:timeevolve}) and (\ref{eq:rho(t+dt)before}) remains Gaussian
and with the same width, but now centered around $\sqrt{2\mu}\mathrm{Tr}%
_{\rho\left(  t\right)  }\left\{  a_{1}+a_{1}^{\dagger}\right\}  \Delta t$.
This means that the variable $\sqrt{\Delta t}X$ is a Gaussian random variable
given by%
\begin{equation}
\sqrt{\Delta t}X=\sqrt{2\mu}\left\langle a_{1}+a_{1}^{\dagger}\right\rangle
\Delta t+\sqrt{2N_{1}+1}dW, \label{eq:sqrt(dt)X}%
\end{equation}
with $dW$ a Wiener increment with $dW^{2}=\Delta t$. These results give us
expressions for the measured current and the effect of the measurement on the
system density matrix.

Since the current is $X/\sqrt{\Delta t}$, the first important result is that
the measured current integrated over time $\Delta t$ is%
\begin{equation}
I\left(  t\right)  \Delta t=\sqrt{2\mu}\left\langle a_{1}+a_{1}^{\dagger
}\right\rangle \Delta t+\sqrt{2N_{1}+1}dW, \label{eq:current}%
\end{equation}
or in differential form
\begin{equation}
I\left(  t\right)  =\sqrt{2\mu}\left\langle a_{1}+a_{1}^{\dagger}\right\rangle
\left(  t\right)  +\sqrt{2N_{1}+1}\xi\left(  t\right)  ,
\end{equation}
where $\xi\left(  t\right)  =dW/dt$ represents white noise with correlations%
\begin{align}
\left\langle \xi\left(  t\right)  \right\rangle  &  =0,\\
\left\langle \xi\left(  t\right)  \xi\left(  t^{\prime}\right)  \right\rangle
&  =\delta\left(  t-t^{\prime}\right)  .
\end{align}
The second result is for the increment of the system density matrix after
evolution through $\Delta t$ and projection by the measurement
(cf.\ \cite{Wth} Eq.\ (4.113)),%
\begin{align}
d\rho^{\mathrm{st}}\left(  t\right)   &  =\langle X|\bar{\rho}(t+\Delta
t)|X\rangle/p(X)\nonumber\\
&  =\sqrt{\Delta t}X\frac{\sqrt{2\mu}}{2N_{1}+1}\left[  \left(  N_{1}%
+1\right)  \left(  a_{1}\rho^{\mathrm{st}}+\rho^{\mathrm{st}}a_{1}^{\dagger
}\right)  \right.  -N_{1}\left(  a_{1}^{\dagger}\rho^{\mathrm{st}}%
+\rho^{\mathrm{st}}a_{1}^{\dagger}\right)  \left.  -\mathrm{Tr}\left\{
a_{1}\rho^{\mathrm{st}}+\rho^{\mathrm{st}}a_{1}^{\dagger}\right\}
\rho^{\mathrm{st}}\right]  +\mathcal{O}\left(  \Delta t\right)  .
\end{align}
Replacing the stochastic variable $X$ by the expression
Eq.\ (\ref{eq:sqrt(dt)X}), and retaining only the $\mathcal{O}\left(
\sqrt{\Delta t}\right)  $ term gives%
\begin{equation}
d\rho^{\mathrm{st}}=\sqrt{\frac{2\mu}{1+2N_{1}}}\left[  (N_{1}+1)(a_{1}%
\rho^{\mathrm{st}}+\rho^{\mathrm{st}}a_{1}^{\dagger})\right.  -\left.
N_{1}(a_{1}^{\dagger}\rho^{\mathrm{st}}+\rho^{\mathrm{st}}a_{1})-\langle
a_{1}+a_{1}^{\dagger}\rangle\rho^{\mathrm{st}}\right]  dW+\mathcal{O}\left(
\Delta t\right)  . \label{eq:dWbeforeAElim}%
\end{equation}
Equation (\ref{eq:dWbeforeAElim}) is the stochastic term that must be added to
the density matrix evolution of equation (\ref{eq:master eqn}) to give us the
stochastic master equation for the density matrix conditioned on the
measurement outcome. Note that the noise term $dW$ appearing in
Eq.\ (\ref{eq:dWbeforeAElim}) is the same as that appearing in
Eq.\ (\ref{eq:current}), so that it is related to the actual current measured
$I(t)$ by Eq. (\ref{eq:current})%
\begin{equation}
dW=\left(  I(t)-\sqrt{2\mu}\left\langle a_{1}+a_{1}^{\dagger}\right\rangle
\right)  \Delta t/\sqrt{1+2N_{1}}.
\end{equation}

\subsection{Adiabatic elimination on the stochastic master equation}

Just as we did for the master equation it is possible to adiabatically
eliminate the ancilla coordinates and find a stochastic master equation for
the system alone. Using the same expansion for the conditioned density matrix
Eq.\ (\ref{eq:tot-densityl}) we can determine stochastic equations for
$\rho_{i}^{\mathrm{st}}$ from Eq.\ (\ref{eq:dWbeforeAElim}). We obtain the set
of differential equations
\begin{align}
d\rho_{0}^{\mathrm{st}}  &  =\sqrt{\frac{2\mu}{2N_{1}+1}}\left\{
(N_{1}+1)\left(  \rho_{1}^{\mathrm{st}}+\rho_{1}^{\mathrm{st}\dagger}\right)
\left\langle \rho_{1}^{\mathrm{st}\dagger}+\rho_{1}^{\mathrm{st}}\right\rangle
\rho_{0}\right\}  dW+\mathcal{O}(dt),\\
d\rho_{1}^{\mathrm{st}}  &  =\sqrt{\frac{2\mu}{2N_{1}+1}}\left\{  \left(
N_{1}+1\right)  \left(  2\rho_{2^{\prime}}^{\mathrm{st}}+\rho_{2}%
^{\mathrm{st}}\right)  -\left\langle \rho_{1}^{\mathrm{st}\dagger}+\rho
_{1}^{\mathrm{st}}\right\rangle \rho_{1}^{\mathrm{st}}\right\}  dW,\\
d\rho_{2}^{\mathrm{st}}  &  =-\sqrt{\frac{2\mu}{2N_{1}+1}}\left\{
\left\langle \rho_{1}^{\mathrm{st}}+\rho_{1}^{\mathrm{st}\dagger}\right\rangle
\rho_{2}^{\mathrm{st}}\right\}  dW.
\end{align}
We have written only the stochastic contributions; the terms proportional to
$dt$ are the same as in the adiabatic elimination on the ordinary master equation.

Now we will do the adiabatic elimination as we did for the deterministic
master equation. As before we wish to set $d\rho_{1}^{\mathrm{st}}$ to zero.
However, since $d\rho_{1}^{\mathrm{st}}$ is stochastically driven this will
not be precisely true even in the steady state. In order to replace $\rho
_{1}^{\mathrm{st}}$ by its mean value at long times it is necessary to
estimate the size of the fluctuations resulting from the $dW$-term. Following
the analysis of Doherty and Jacobs \cite{DJ99} we integrate the stochastic
term over the decay time of the ancilla oscillator and compare its root mean
square magnitude with the deterministic terms. Consider the full equation for
$d\rho_{1}$\
\begin{align}
d\rho_{1}^{\mathrm{st}}=  &  -i\lambda_{01}\alpha\left[  a_{0}^{\dagger}%
a_{0}\rho_{0}^{\mathrm{st}}-\frac{N_{1}}{N_{1}+1}\rho_{0}^{\mathrm{st}}%
a_{0}^{\dagger}a_{0}\right]  dt\nonumber\\
&  -\kappa\rho_{1}dt+\sqrt{\frac{2\mu}{2N_{1}+1}}\left\{  \left(
N_{1}+1\right)  \left(  2\rho_{2^{\prime}}^{\mathrm{st}}+\rho_{2}\right)
\right.  \left.  -\left\langle \rho_{1}^{\mathrm{st}}+\rho_{1}^{\mathrm{st}%
\dagger}\right\rangle \rho_{1}^{\mathrm{st}}\right\}  dW+\kappa\mathcal{O}%
\left(  \epsilon^{3}\right)  dt.
\end{align}
We integrate this over a time $\Delta t\sim1/\kappa$ and use the fact that the
mean values of $\rho_{0},\rho_{1}$ must be slowly varying over this time to
obtain
\begin{align}
\Delta\rho_{1}  &  =\int_{0}^{\Delta t}d\rho_{1}\\
\simeq &  -i\lambda_{01}\alpha\left[  a_{0}^{\dagger}a_{0}\rho_{0}-\frac
{N_{1}}{N_{1}+1}\rho_{0}a_{0}^{\dagger}a_{0}\right]  \Delta t-\kappa\rho
_{1}\Delta t\nonumber\\
&  +\sqrt{\frac{2\mu}{2N_{1}+1}}\left\{  \left(  N_{1}+1\right)  \left(
2\rho_{2^{\prime}}+\rho_{2}\right)  \right.  \left.  -\left\langle \rho
_{1}+\rho_{1^{\dagger}}\right\rangle \rho_{1}\right\}  \int_{0}^{\Delta
t}dW(t^{\prime})+\mathcal{O}\left(  \epsilon^{3}\right)  .
\label{eq:stochchange}%
\end{align}
The random number $\Delta W=\int_{0}^{\Delta t}dW(t^{\prime})$ is Gaussian
distributed with mean zero and variance $\Delta t\sim1/\kappa$ \cite{Ga2},
thus the root mean square size of $\Delta W$ is $1/\sqrt{\kappa}$. As a result
the stochastic term in the update of $\rho_{1}$ scales like $\epsilon^{5/2}$
and is negligible in comparison to the deterministic terms which scale like
$\epsilon$. As a result Eq.\ (\ref{eq:rho1}) holds exactly as before. Using
Eq.\ (\ref{eq:Tr rho}) and $i\alpha=-i\alpha^{\ast}=|\alpha|$ we finally
obtain the stochastic master equation (SME) for the system%
\begin{align}
d\rho_{\mathrm{s}}  &  =-\left\{  \frac{\lambda_{01}^{2}\left\vert
\alpha\right\vert ^{2}(2N_{1}+1)}{\kappa}\left[  a_{0}^{\dagger}a_{0},\left[
a_{0}^{\dagger}a_{0},\rho_{\mathrm{s}}\right]  \right]  \right\}  dt-i\left\{
\omega_{0}+\lambda_{01}(\left\vert \alpha\right\vert ^{2}+N_{1})\right\}
\left[  a_{0}^{\dagger}a_{0},\rho_{\mathrm{s}}\right]  dt\nonumber\\
&  +\nu\left(  N_{0}+1\right)  \left(  2a_{0}\rho_{\mathrm{s}}a_{0}^{\dagger
}-a_{0}^{\dagger}a_{0}\rho_{\mathrm{s}}-\rho_{\mathrm{s}}a_{0}^{\dagger}%
a_{0}\right)  dt+\nu N_{0}\left(  2a_{0}^{\dagger}\rho_{\mathrm{s}}a_{0}%
-\rho_{\mathrm{s}}a_{0}a_{0}^{\dagger}-a_{0}a_{0}^{\dagger}\rho_{\mathrm{s}%
}\right)  dt\nonumber\\
&  -\sqrt{2k}\left[  a_{0}^{\dagger}a_{0}\rho_{\mathrm{s}}+\rho_{\mathrm{s}%
}a_{0}^{\dagger}a_{0}-2\langle a_{0}^{\dagger}a_{0}\rangle\rho_{\mathrm{s}%
}\right]  dW, \label{eq:redstochmastereqn}%
\end{align}
where%
\begin{equation}
k\equiv\mu\lambda_{01}^{2}|\alpha|^{2}/\left(  2N_{1}+1\right)  \kappa^{2}.
\label{eq:k}%
\end{equation}
Again the noise $dW$ is related to the measured current, which using
Eq.\ (\ref{eq:rho1}) can be written in terms of the system phonon number
\begin{equation}
I(t)\Delta t=\sqrt{2N_{1}+1}\left(  2\sqrt{2k}\left\langle a_{0}^{\dagger
}a_{0}\right\rangle \Delta t+dW\right)  . \label{eq:scaled-I(t)}%
\end{equation}

\section{Results\label{SubSec_competition}}

To further understand the consequences of the stochastic density matrix
equation Eq.\ (\ref{eq:redstochmastereqn}) we consider a case in which the
initial state is a mixture of number states, $\rho_{s}=\sum_{n}p_{n}%
|n\rangle\langle n|$. (A thermal state is an example of such a state.) The
solution of the stochastic master equation, Eq.\ (\ref{eq:redstochmastereqn}),
remains a mixture of number states if the initial state is a mixture of number
states. For such an initial state the stochastic master equation can be
reduced to an equation for the weights $p_{n}$, which takes the form%
\begin{align}
dp_{n}  &  =-2\nu(N_{0}+1)[np_{n}-(n+1)p_{n+1}]dt\nonumber\\
&  -2\nu N_{0}[(n+1)p_{n}-np_{n-1}]dt-2\sqrt{2k}\left(  n-\sum_{n^{\prime}%
}n^{\prime}p_{n^{\prime}}\right)  p_{n}dW. \label{eq:numdist}%
\end{align}
Since mixtures of number states are invariant under changes of phase and the
number states are eigenstates of the Hamiltonian, neither the phase diffusion
term nor the Hamiltonian terms in the stochastic master equation contribute to
the evolution of the phonon number distribution. It can also be shown that
this system of equations also describes the evolution of the phonon number
distribution $p_{n}=\langle n|\rho|n\rangle$ for an arbitrary (not necessarily
diagonal) initial state $\rho(0)$.

Equation (\ref{eq:numdist}) is our central result for analyzing the behavior
of the measurement protocol. The first two terms of Eq.\ (\ref{eq:numdist})
containing $dt$ describe emission into and absorption from the thermal bath
coupled to the system. We will see that the second, stochastic, term tends to
concentrate the distribution $p(n)$ onto a single value of $n$. The two
effects are characterized by the two time scales, $\nu^{-1}$ and $k^{-1}$, and
the resulting behavior depends on the ratio of these two times. We will
discuss the competition by calculating the occupation number of the system,
$\left\langle a_{0}^{\dagger}a_{0}\right\rangle \left(  t\right)  $, which is
given in terms of $p_{n}$ from numerical simulations of Eq.\ (\ref{eq:numdist}%
) by%
\begin{equation}
\left\langle a_{0}^{\dagger}a_{0}\right\rangle \left(  t\right)  =\sum
_{n}np_{n}\left(  t\right)  . \label{eq:a0a0}%
\end{equation}

\begin{figure}[ptb]
\begin{center}
\includegraphics[width=3.1in]{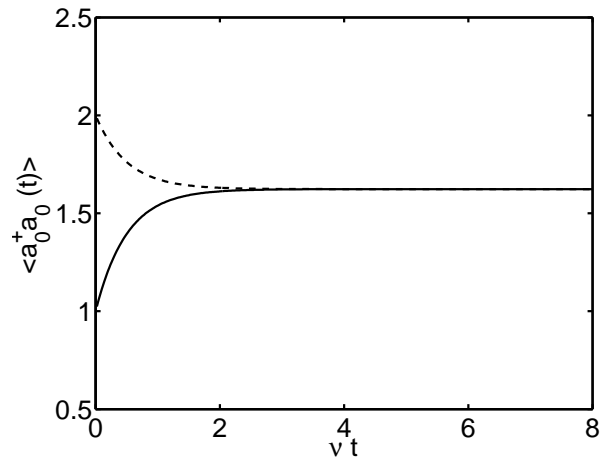}
\end{center}
\caption{Plot of the solution of Eq.\ (\ref{eq:numdist}) without
the stochastic component, $k=0$, with the initial state
$\left\vert 1\right\rangle
$ (solid line) and $\left\vert 2\right\rangle $ (dashed line).}%
\label{fig:pneqdtn}%
\end{figure}

Firstly we turn off the stochastic component and consider the solutions given
by the deterministic part of Eq.\ (\ref{eq:numdist}). Figure \ref{fig:pneqdtn}
is a plot for $k=0$ starting from two different initial states $\left\vert
1\right\rangle $ and $\left\vert 2\right\rangle $ and for a bath temperature
corresponding to an average occupation number $N_{0}=1.62$. The plot shows
that the deterministic terms in Eq.\ (\ref{eq:numdist}) drive the system
towards a mixed (thermal) state, so that the ensemble average of $\left\langle
a_{0}^{\dagger}a_{0}\right\rangle (t)$ gradually reaches the thermal average
at the bath temperature. This is true regardless of the initial state. Note
that the deterministic part of Eq.\ (\ref{eq:numdist}) also describes the
average over all measurement outcomes even for nonzero $k,$since the
stochastic term averages to zero. We can define the characteristic time the
system resides in a given number state, which we call the dwell time
$t_{\mathrm{dwell}}$, as the reciprocal of the initial transition rate given
by Eq.\ (\ref{eq:numdist}) with $k=0$
\begin{equation}
t_{\mathrm{dwell}}=\left.  \frac{dp_{n}}{dt}\right\vert _{p_{n}(0)=1}=\frac
{1}{2\nu\left[  N_{0}(n+1)+\left(  N_{0}+1\right)  n\right]  }.
\label{eq:dwelltime}%
\end{equation}
Note that the dwell time depends on the initial state $n$, and also on the
temperature of the bath through $N_{0}$.

\begin{figure}[ptb]
\begin{center}
\includegraphics[width=3.1 in]{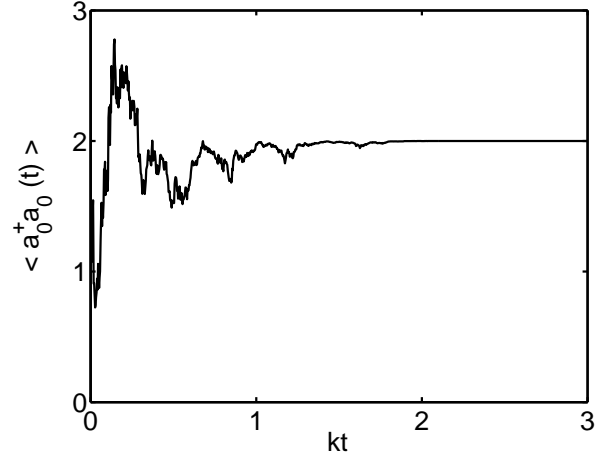}
\end{center}
\caption{A plot of a solution to Eq.\ (\ref{eq:numdist}) with
$\nu=0$\ with an
initial state that is thermal.}%
\label{fig:pneqdwNonly}%
\end{figure}

\begin{figure}[ptb]
\begin{center}
\includegraphics[width=3.1 in]{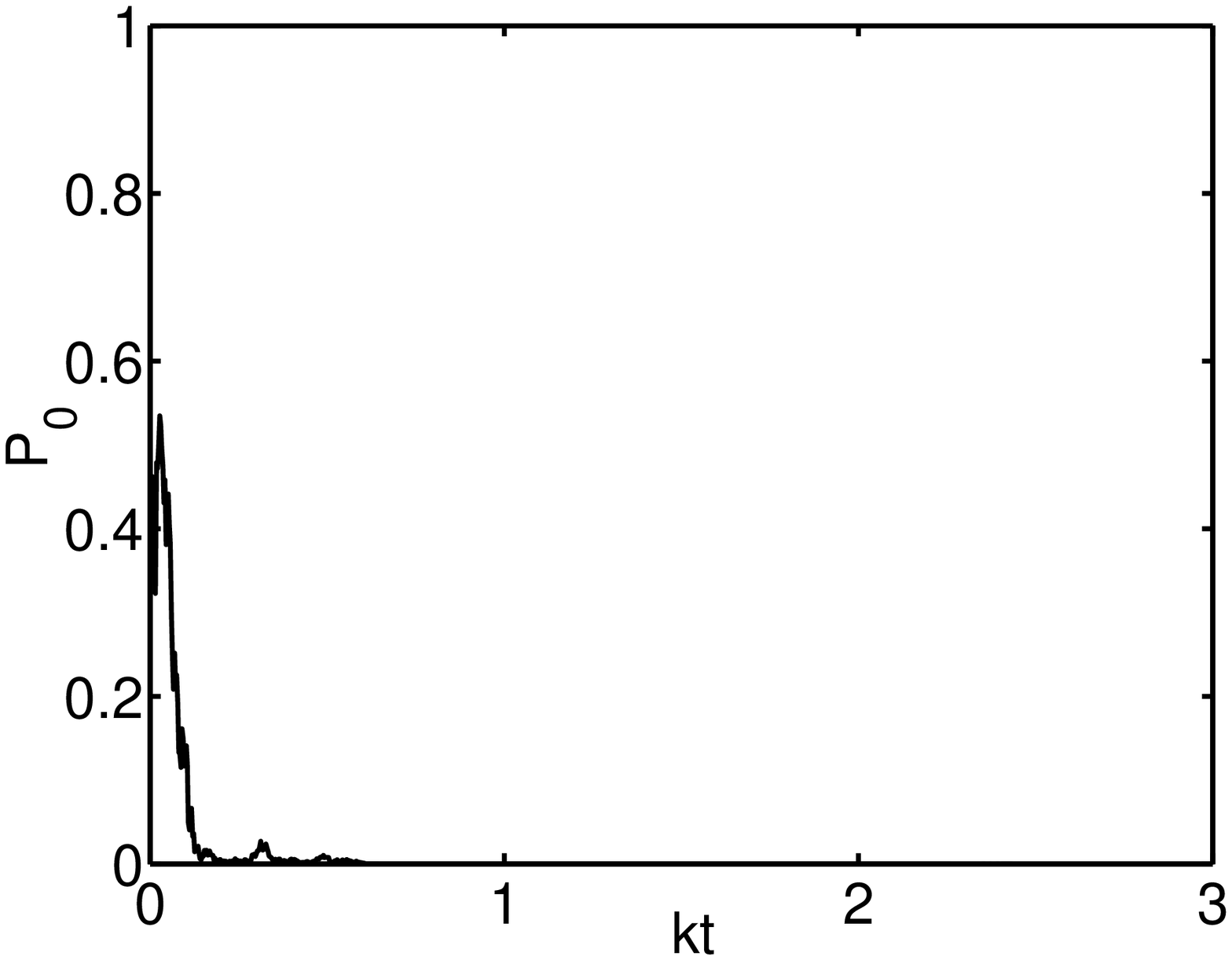}
\includegraphics[width=3.1 in]{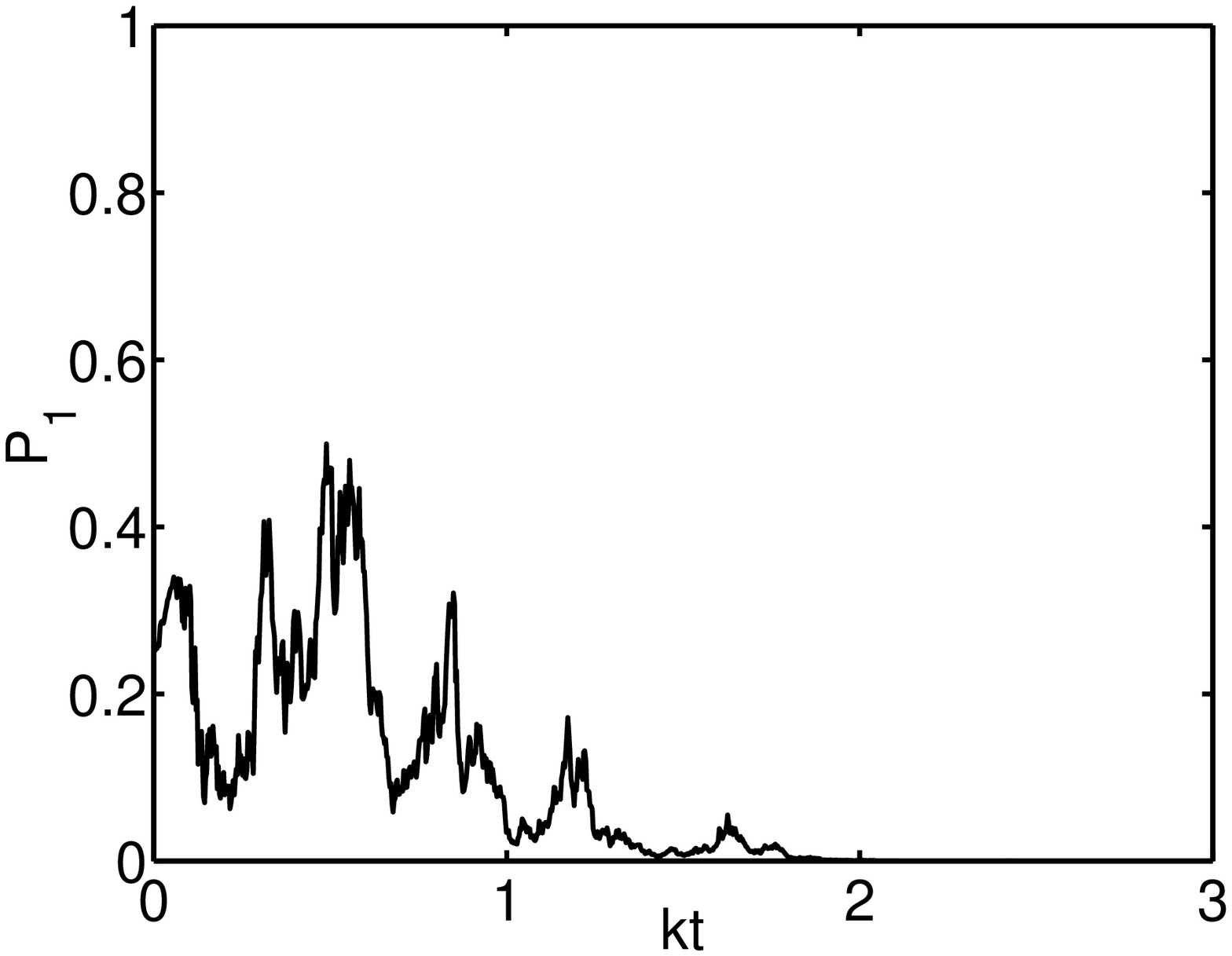}
\includegraphics[width=3.1 in]{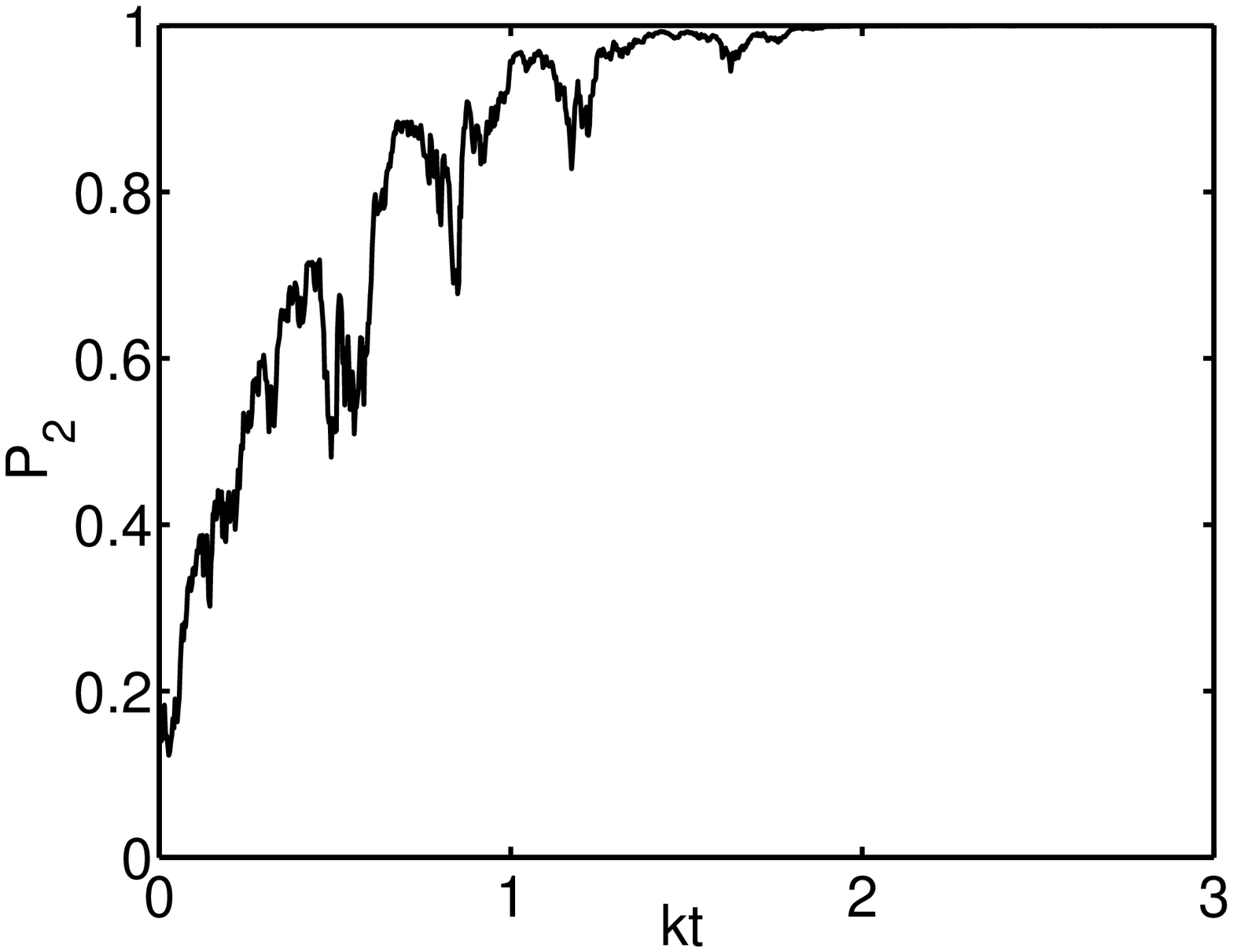}
\includegraphics[width=3.1 in]{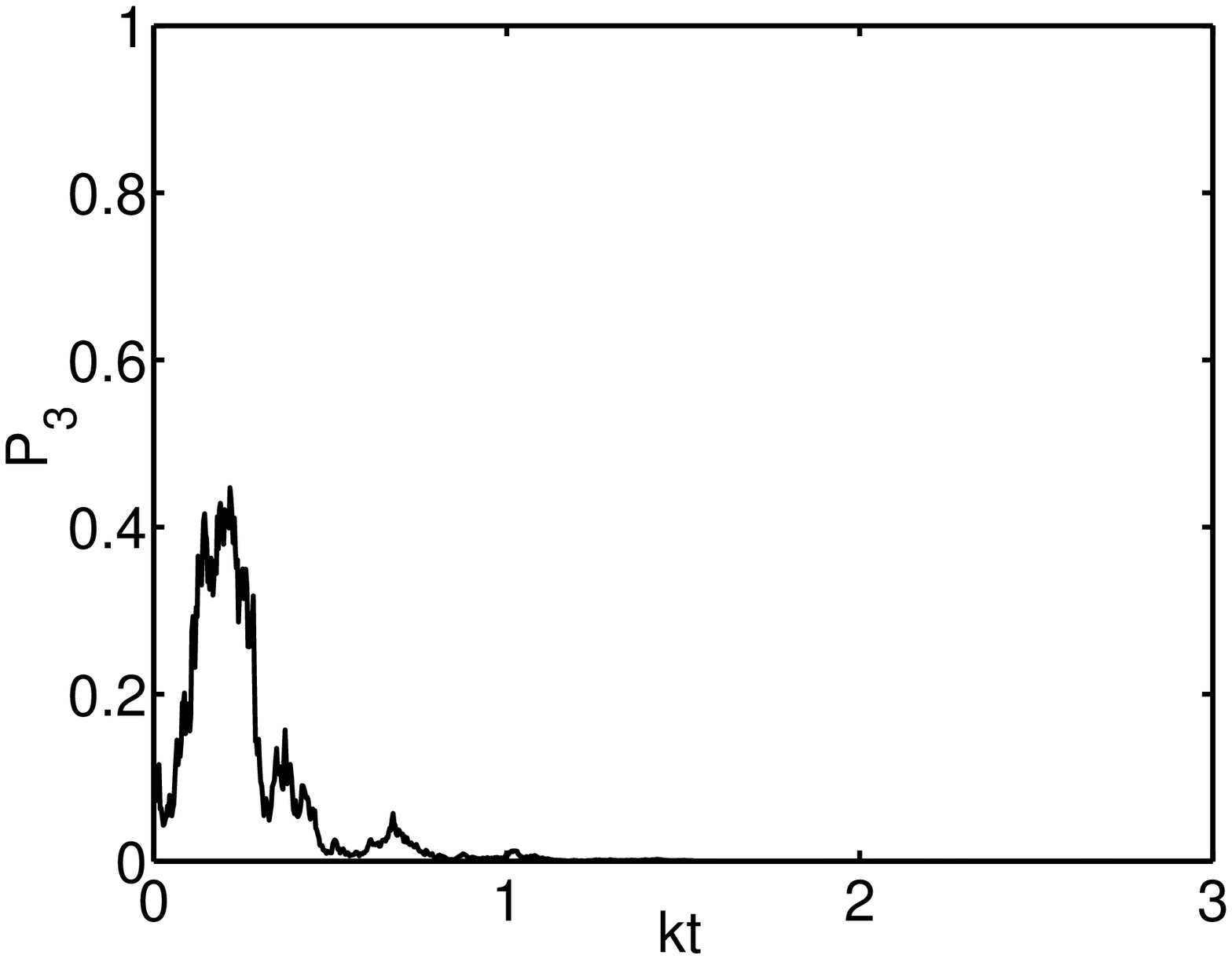}
\end{center}
\caption{Plot of $p_{n}(t)$ for a simulation of Eq.\
(\ref{eq:numdist}) with $\nu=0$ for the states $\left\vert
0\right\rangle ,\left\vert 1\right\rangle ,\left\vert
2\right\rangle ,\left\vert 3\right\rangle $. The initial state is
a thermal state with the average occupation number $1.63$. The
figure is for
the same simulation as in Fig.\ \ref{fig:pneqdwNonly}.}%
\label{fig:pneqdwn}%
\end{figure}

We turn now to the dynamics resulting from the measurement process
in the absence of coupling to the thermal bath. Figure\
\ref{fig:pneqdwNonly} shows results for $\left\langle
a_{0}^{\dagger}a_{0}\right\rangle \left(  t\right) $ for a
simulation of Eq.\ (\ref{eq:numdist}) with $\nu=0$ and an initial
condition of a thermal state. Figure\ \ref{fig:pneqdwn} shows the
individual probabilities $p_{n}$ for $n=0,1,2,3$ for the same
simulation. All number states are present initially, but
eventually the system is projected onto state $\left\vert
1\right\rangle $ in this simulation. In other runs, with different
random numbers for the stochastic term, different final states
result, as expected. The plots show that the stochastic term tends
to project the system state onto a pure number state on a time
scale of order $k^{-1}$. We call this time the collapse time
$t_{\text{\textrm{coll}}}$. Since no coupling to the thermal bath
is present in these simulations, once projected onto a number
state, the state is stationary. The collapse onto a number state
can actually be shown analytically using the solution of the
system of equations (\ref{eq:numdist}) due to Jacobs and
Knight~\cite{Ja1998}.

\begin{figure}[ptb]
\begin{center}
\includegraphics[width=3.1 in]{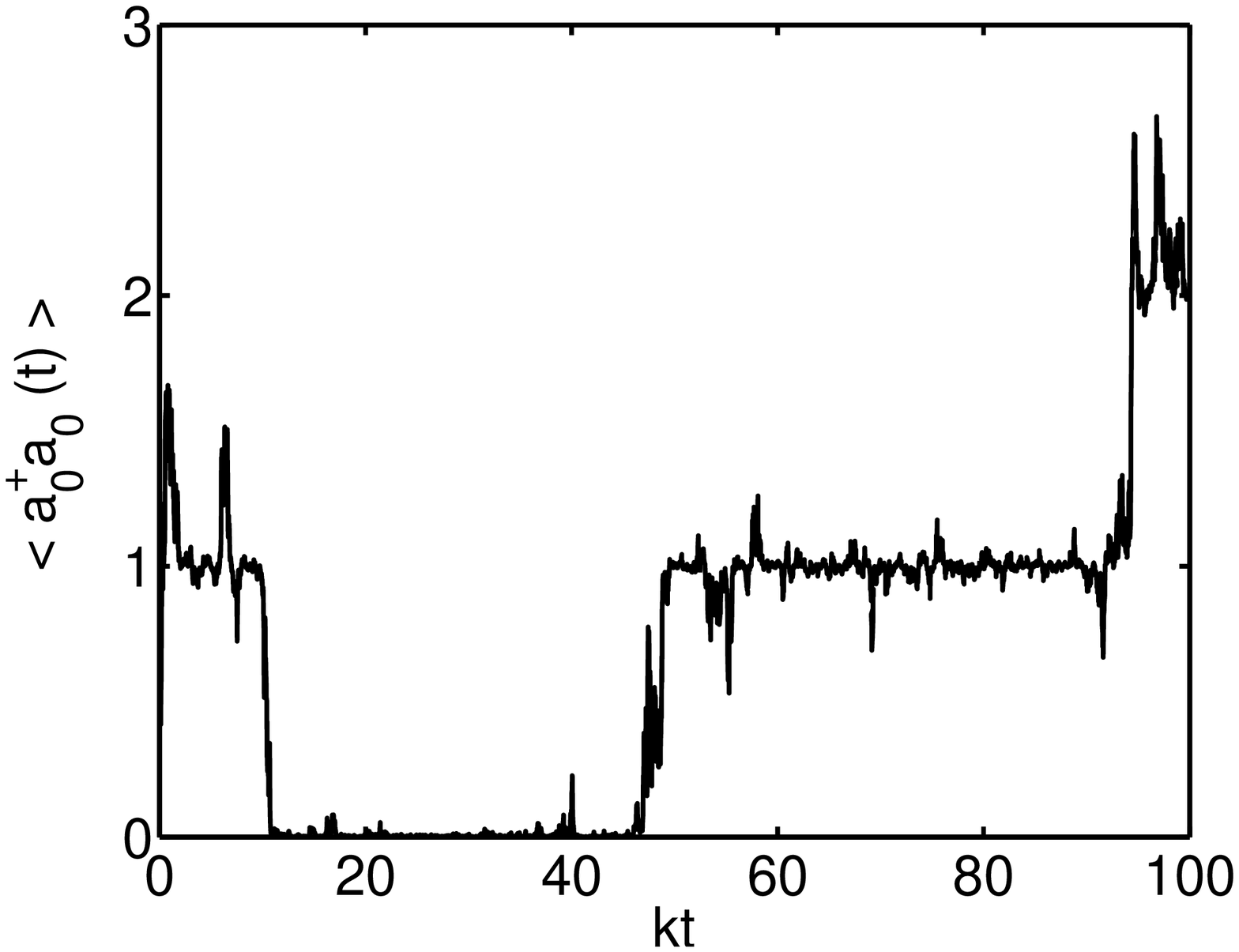}
\includegraphics[width=3.0 in]{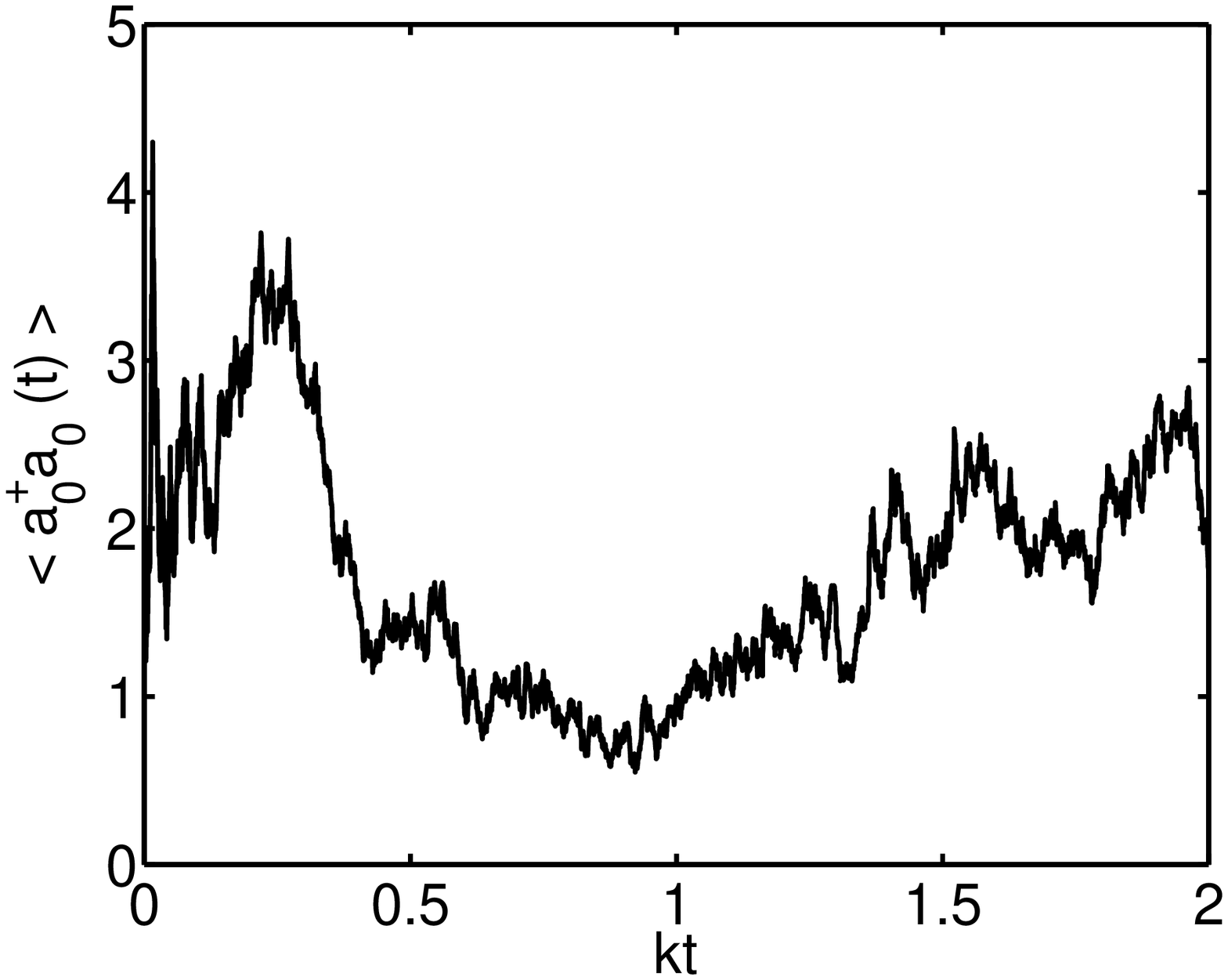}
\end{center}
\caption{The evolution of the phonon number $\left\langle a_{0}^{\dagger}%
a_{0}\right\rangle (t)$ given by Eq.\ (\ref{eq:numdist}) using
$N_{0}=1.62$
and $k/\nu=250$ (first panel) and $k/\nu=5$ (second panel).}%
\label{fig:k5np02}%
\end{figure}

\begin{figure}[ptb]
\begin{center}
\includegraphics[width=3.1 in]{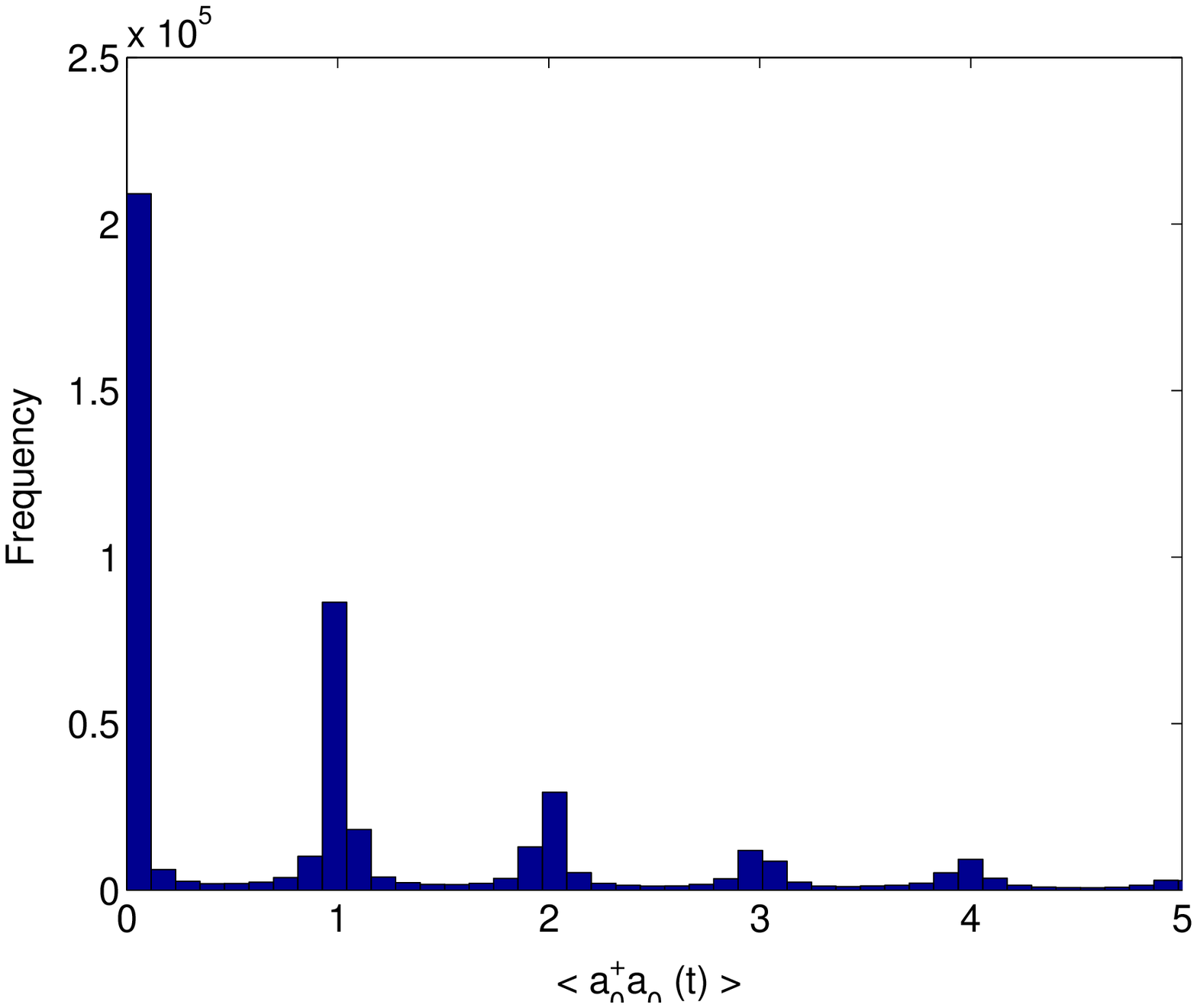}
\includegraphics[width=3.0 in]{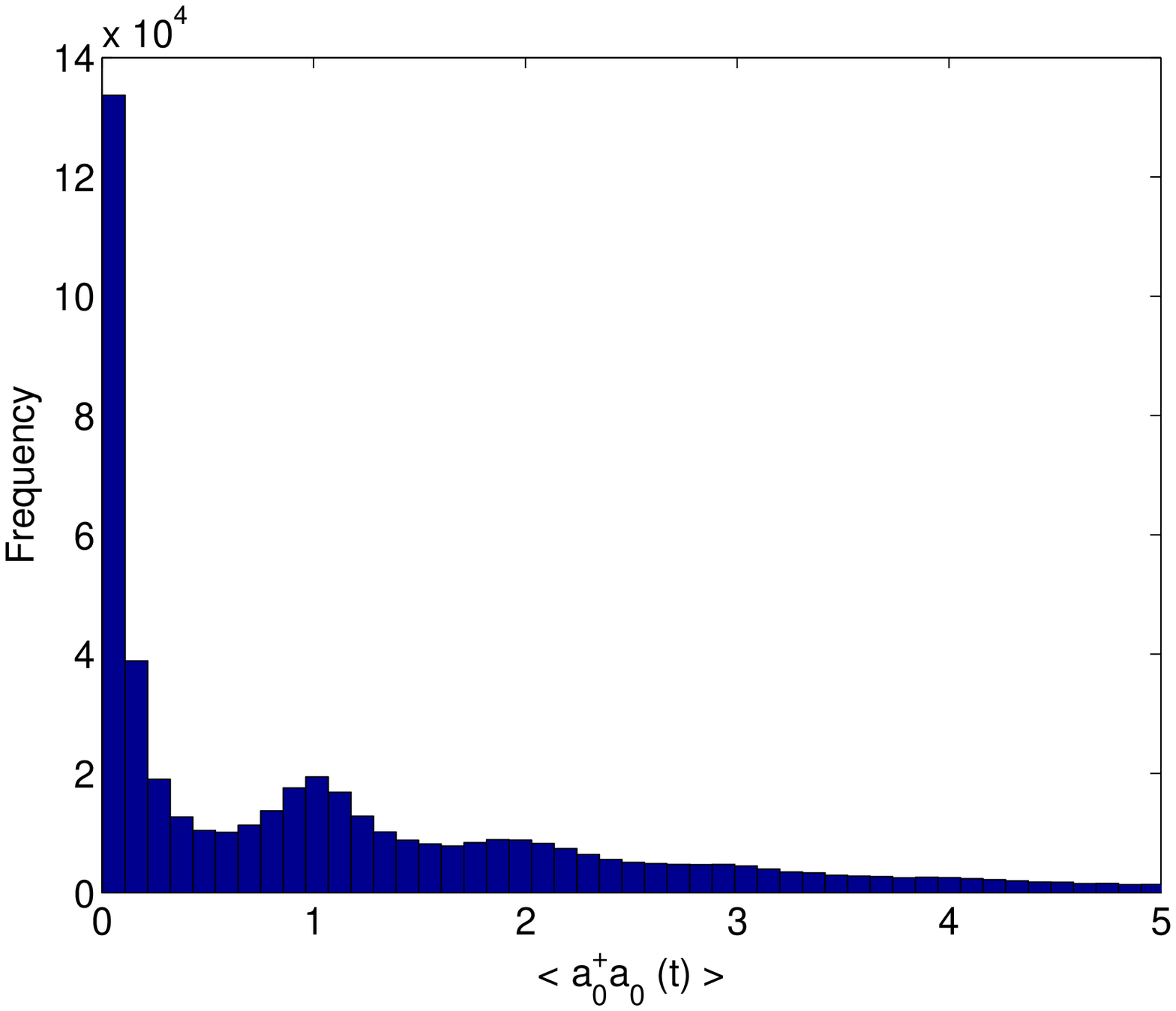}
\end{center}
\caption{Histogram of $\left\langle
a_{0}^{\dagger}a_{0}\right\rangle \left( t\right)  $ for a long
simulation with $k/\nu=150$ and $N_{0}=1.62$ (first
panel) and $k/\nu=15$ (second panel).}%
\label{fig:histjumpsnp02}%
\end{figure}

\begin{figure}[ptb]
\begin{center}
\includegraphics[width=3.1 in]{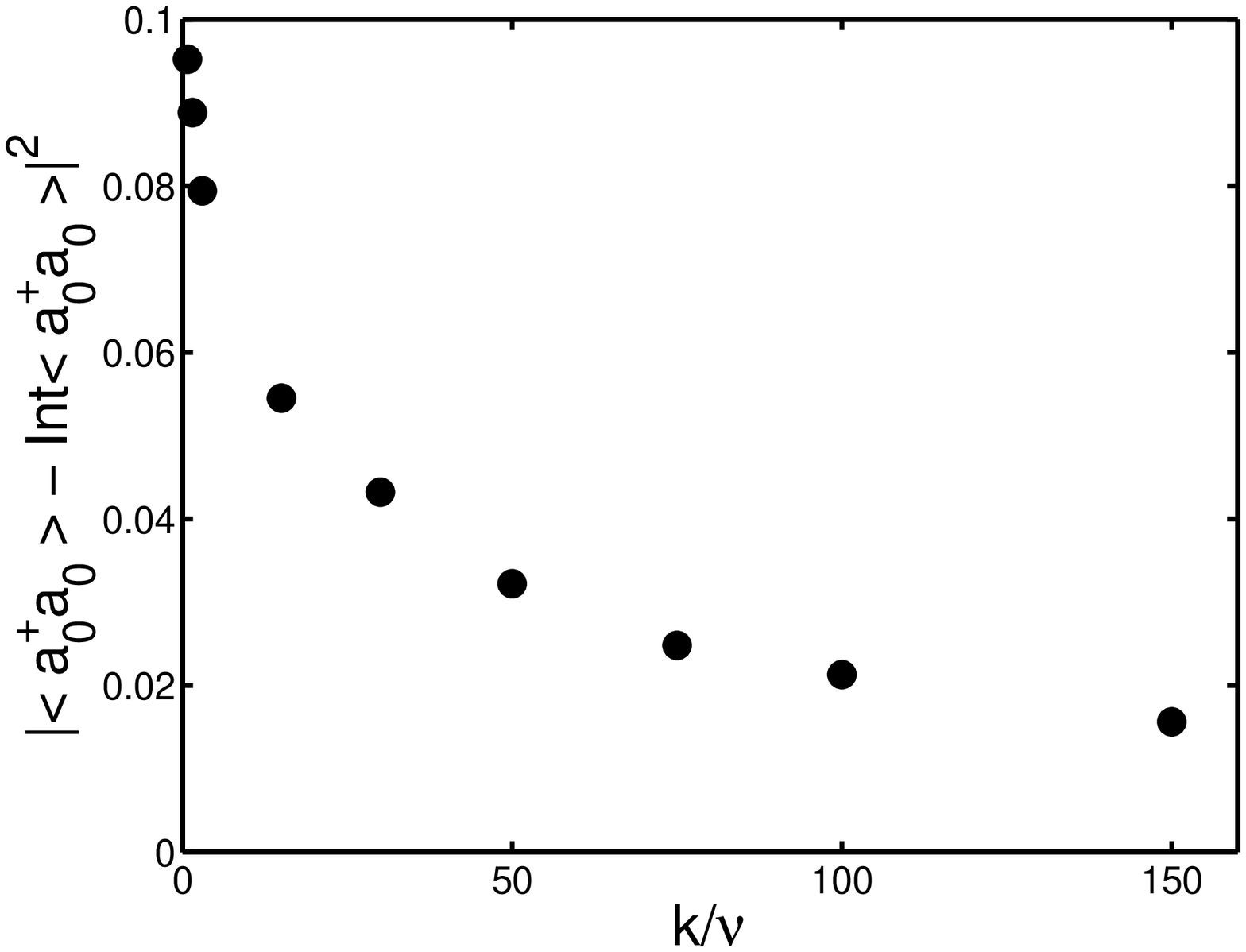}
\end{center}
\caption{The\ deviation of $\left\langle
a_{0}^{\dagger}a_{0}\right\rangle \left(  t\right)  $ from
integral values as a function of $k/\nu$. The deviation is defined
as the average of ${\textstyle}\left\vert \left\langle
a_{0}^{\dagger}a_{0}\right\rangle \left(  t\right)
-\mathrm{Int}\left\langle
a_{0}^{\dagger}a_{0}\right\rangle \left(  t\right)  \right\vert ^{2}$.}%
\label{fig:n-intn2}%
\end{figure}

For the phonon number $\left\langle a_{0}^{\dagger}a_{0}\right\rangle \left(
t\right)  $ to take on discrete values with both the thermalization by the
coupling to the bath and the projection by the measurement process present, we
need $t_{\mathrm{dwell}}\gtrsim t_{\mathrm{coll}}$. This is illustrated in
Figs.\ \ref{fig:k5np02}, which show results for the cases $t_{\mathrm{dwell}%
}\gg t_{\mathrm{coll}}$ and the $t_{\mathrm{dwell}}\ll t_{\mathrm{coll}}$ with
the fixed value of $N_{0}=1.62$. We use values of $k/\nu=250$ giving
$t_{\mathrm{dwell}}/t_{\mathrm{coll}}=153$ for state $\left\vert
0\right\rangle $ and $t_{\mathrm{dwell}}/t_{\mathrm{coll}}=42.4$ for state
$\left\vert 1\right\rangle $, and $k/\nu=5$ giving $t_{\mathrm{dwell}%
}/t_{\mathrm{coll}}=3.06$ for state $\left\vert 0\right\rangle $ and
$t_{\mathrm{dwell}}/t_{\mathrm{coll}}=0.85$ for state $\left\vert
1\right\rangle $. The jumps in the occupation number are clearly evident in
the former case, but are not seen in the latter case. The discreteness in the
phonon number is shown more clearly by plotting histograms of $\left\langle
a_{0}^{\dagger}a_{0}\right\rangle (t)$, Figs.\ \ref{fig:histjumpsnp02}, again
using a fixed value of $N_{0}=1.62$ but with different values of $k/\nu$ equal
to $150$ and $15$. (A bin width $\Delta\left\langle a_{0}^{\dagger}%
a_{0}\right\rangle =0.1$ is used.) The clustering of the $\left\langle
a_{0}^{\dagger}a_{0}\right\rangle (t)$ values around integral values is
clearly evident for $k/\nu=150$, is still identifiable for $k/\nu=15$, and
completely absent for $k/\nu=3$. The increasing sharpness of the jumps with
larger $k/\nu$ can be seen in a more quantitative manner by plotting the
standard deviation of the phonon number from integer values, the time and
ensemble average of $\left\vert \left\langle a_{0}^{\dagger}a_{0}\right\rangle
(t)-\mathrm{Int}\left\langle a_{0}^{\dagger}a_{0}\right\rangle (t)\right\vert
^{2}$, as a function of $k/\nu$ (see Fig.\ \ref{fig:n-intn2}).

\begin{figure}[ptb]
\begin{center}
\includegraphics[width=3.1in]{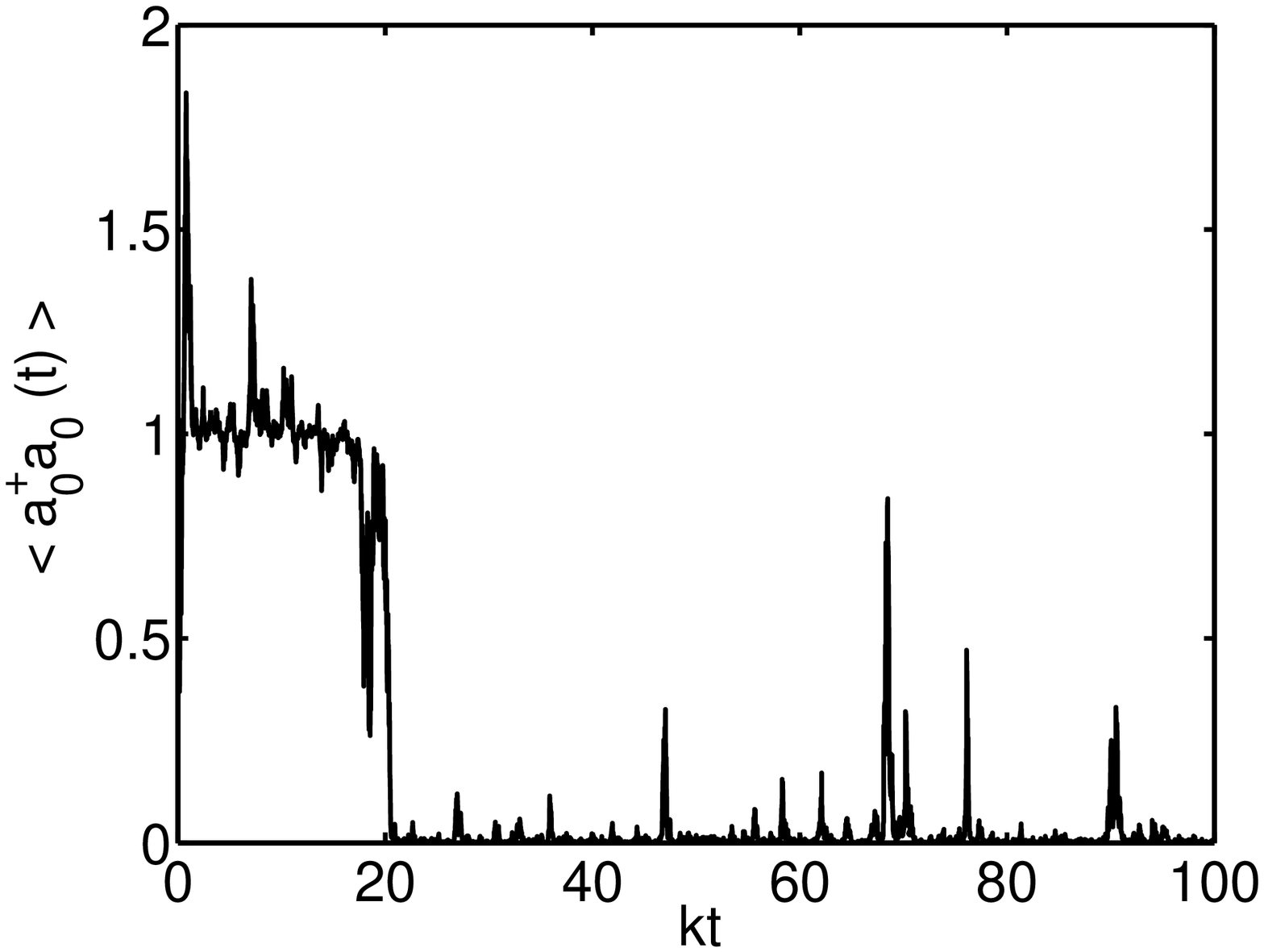}
\includegraphics[width=3.0in]{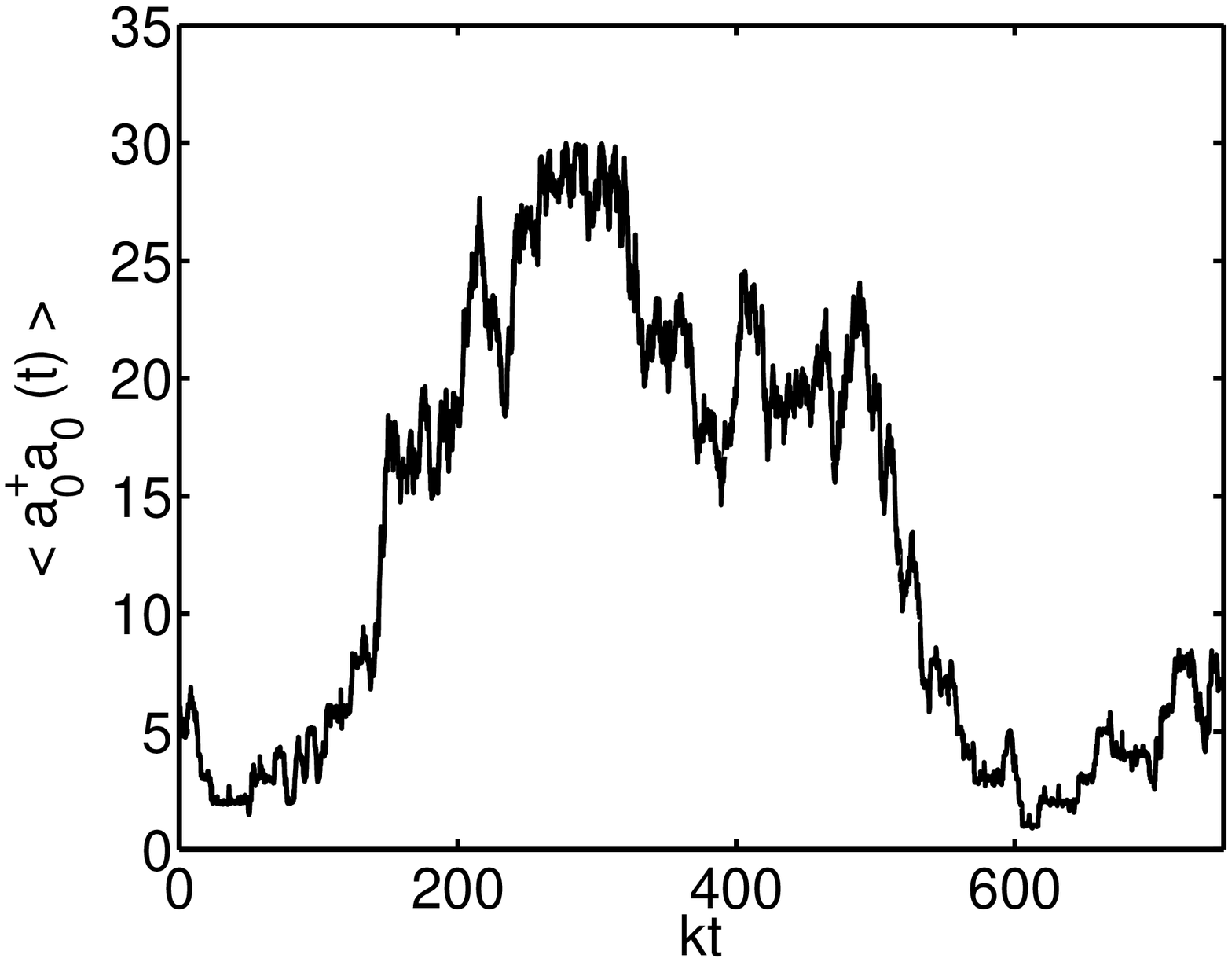}
\end{center}
\caption{Evolution of $\left\langle
a_{0}^{\dagger}a_{0}\right\rangle (t)$ at a temperature
corresponding to $N_{0}=1.62$ (first panel) and $N_{0}=20$ (second
panel), with $\nu N_{0}/k=0.0108$ and from an initial state
$\left\vert 2\right\rangle $.}%
\label{fig:lowT}%
\end{figure}

Since $t_{\mathrm{dwell}}$ is dependent on the temperature, the condition
$t_{\mathrm{dwell}}\gtrsim t_{\mathrm{coll}}$ effectively places a limit on
the temperature of the system oscillator even for large $k$. Setting $n=N_{0}$
in Eq.\ (\ref{eq:dwelltime}) for $t_{\mathrm{dwell}}$, this inequality gives
the condition on the temperature for jumps in the number to be seen
\begin{equation}
\nu N_{0}(N_{0}+1)/2k\lesssim1.
\end{equation}
In order to keep the same resolution for observing clear jumps as
at low temperature, $k/\nu$ must be increased as temperature
increases. This is not an easy task for the experimenters: for
example, an oscillator with $1$ GHz resonant frequency, which is
the highest frequency currently reported for a mesoscopic
oscillator \cite{HZMR03}, at $T=0.1$ K the average occupation
number is $N_{0}=1.62$. When the temperature is raised to $T=1$ K,
the value rises to $N_{0}=20$. Thus if we demand the same
resolution for jumping in both cases, the sensitivity of the
measurement at the higher temperature must be increased by a large
factor. This is illustrated by Figs.\ \ref{fig:lowT} which show
$\left\langle a_{0}^{\dagger}a_{0}\right\rangle (t)$ over time for
different temperatures corresponding to $N_{0}=1.62$ for
$k/\nu=150$, and $N_{0}=20$ for $k$/$\nu=1850$. The product $\nu
N_{0}/k$ has been kept constant at $0.0108$ in order to provide
the same resolution for the jumps. Also notice from Eq.\
(\ref{eq:dwelltime}) that $t_{\mathrm{dwell}}$ decreases with the
system state $n$ making it difficult to recognize the discrete
jumps when the system state is at higher $n$.

\begin{figure}[ptb]
\begin{center}
\includegraphics[width=3.0in]{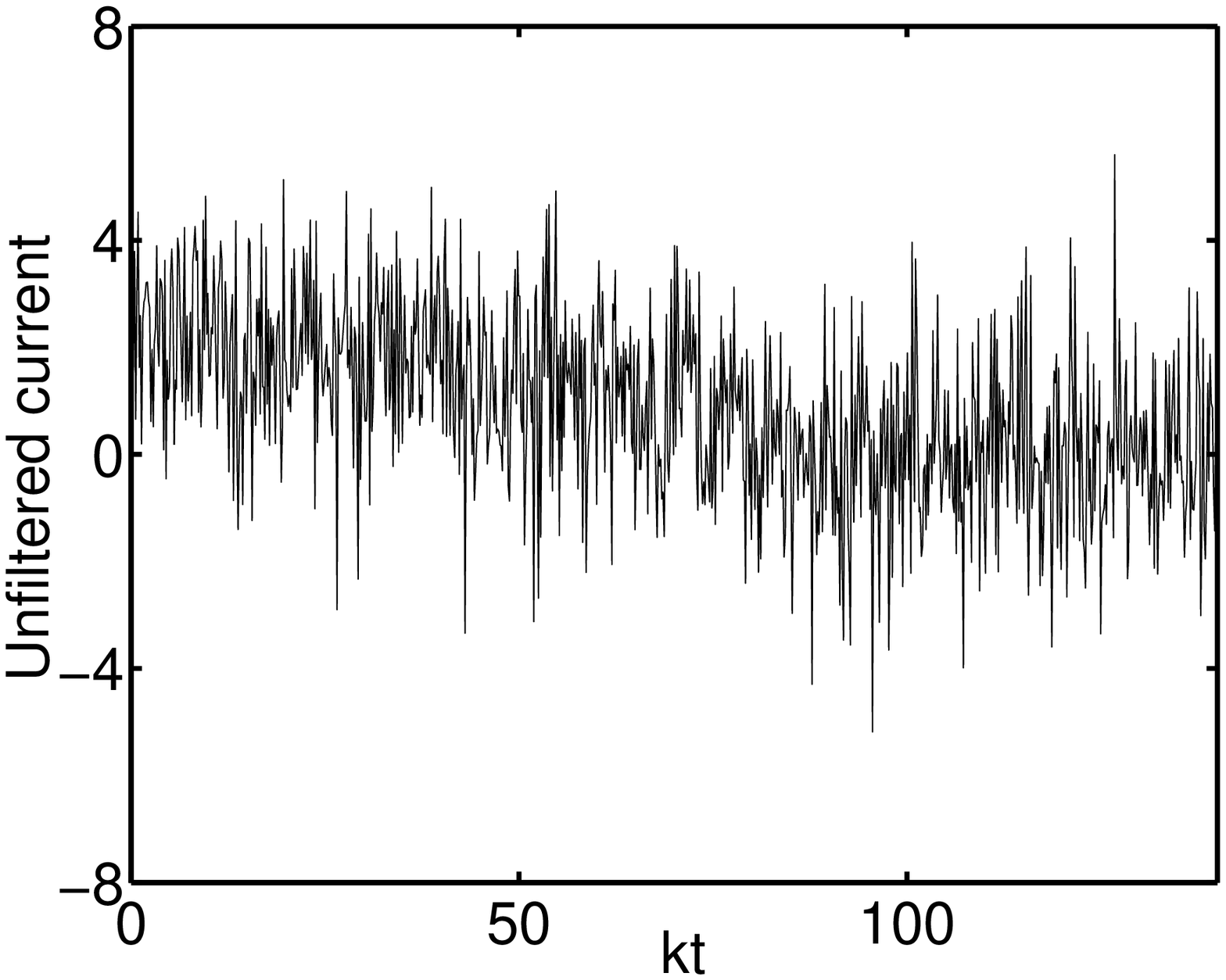}
\includegraphics[width=3.0in]{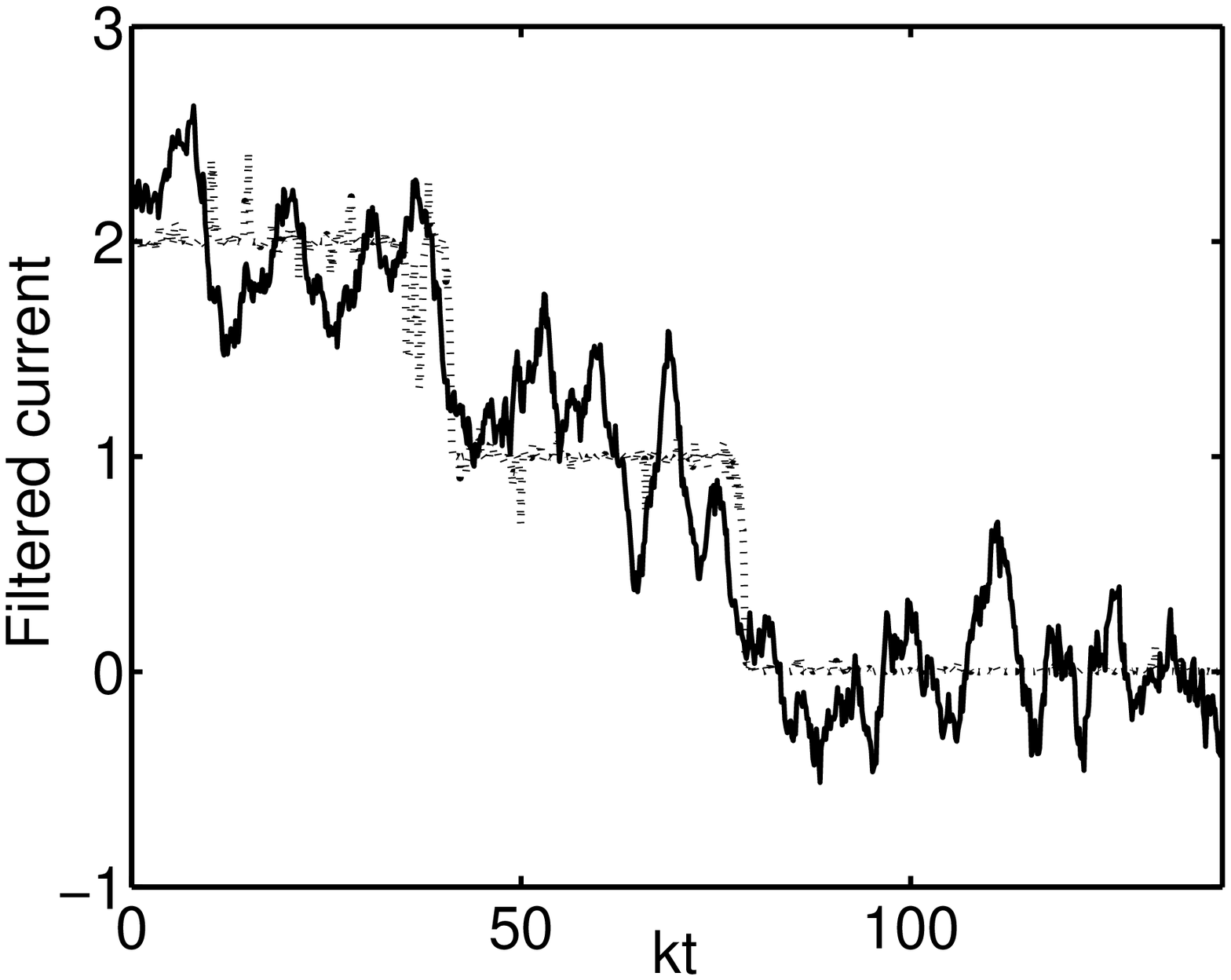}
\includegraphics[width=3.0in]{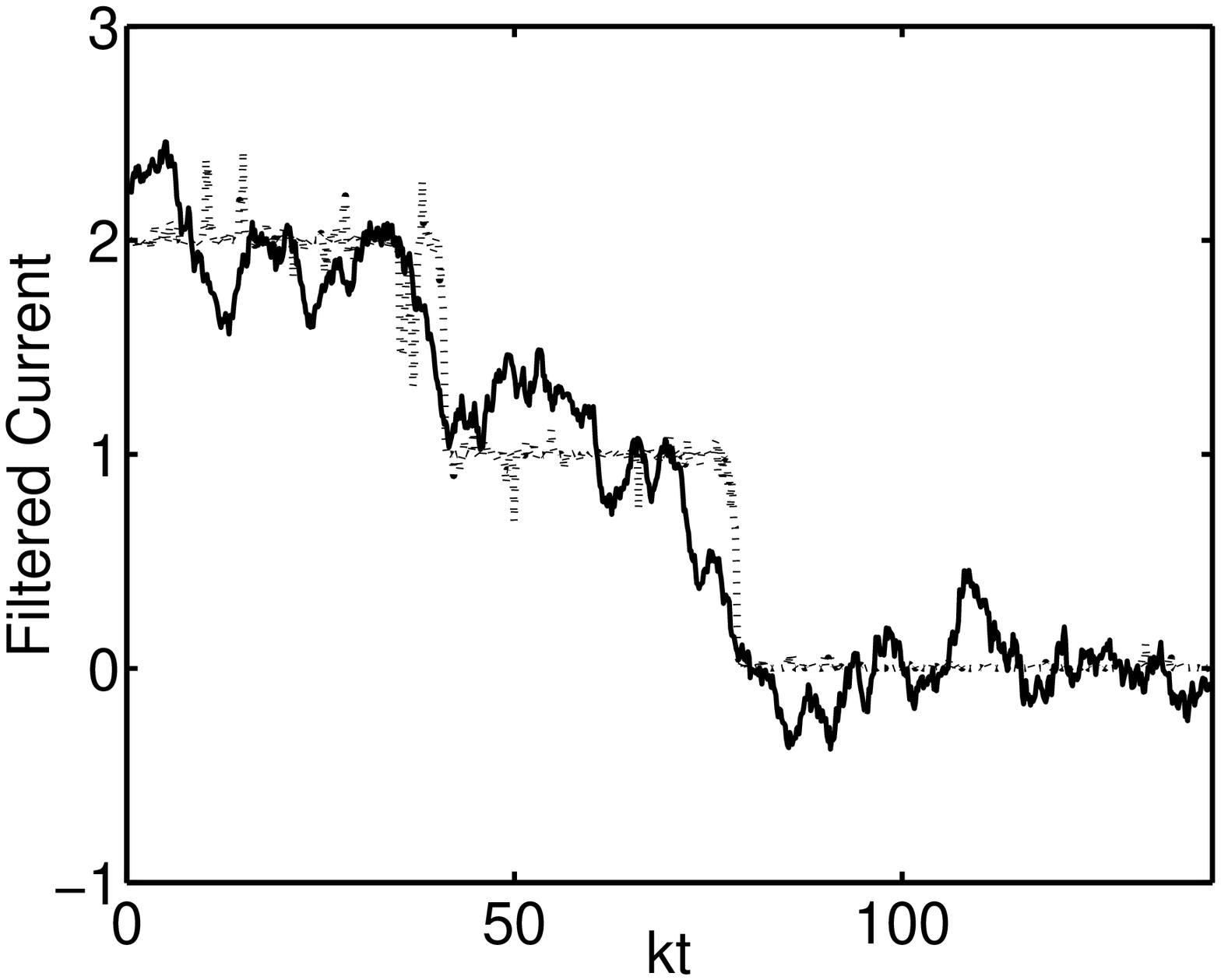}
\includegraphics[width=3.0in]{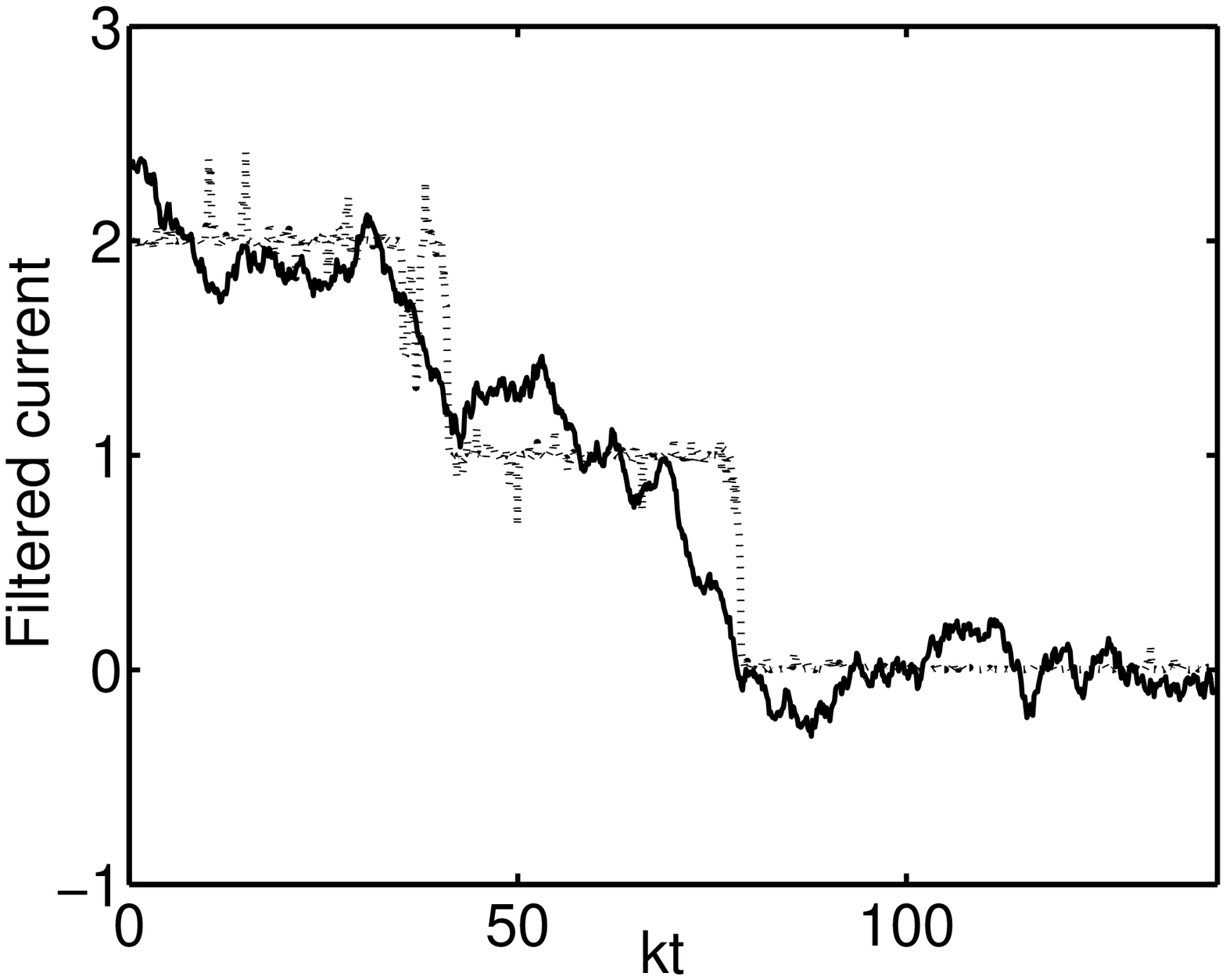}
\end{center}
\caption{Filtered current using a running average over various
window sizes, for parameters $k/\nu=250$ and $N_{0}=1.62$. The
dotted line is $\left\langle a_{0}^{\dagger}a_{0}\right\rangle
(t)$ given by the stochastic density matrix. The current was first
averaged over a time interval of $k\Delta t=0.15$. Upper left:
current observed; upper right: window size $k\Delta t=$ $4.5$;
lower
left: window size $k\Delta t=7.5$; lower right: window size $k\Delta t=10.5$.}%
\label{fig:filter}%
\end{figure}

We have so far considered the possibility of observing discrete occupation
numbers in terms of the behavior of the variable $\left\langle a_{0}^{\dagger
}a_{0}\right\rangle (t)$. In actual practice the occupation number must be
inferred from the measured current $I(t)$, and is obscured by the noise in
this variable. A simple scheme to reduce the effect of the noise is to average
the signal over a sliding window. We can define the measurement time
$t_{\mathrm{m}}$ as the averaging time required to give unit signal to noise
ratio. Thus we equate the signal $S$, given by averaging the current for unit
phonon number $\left\langle a_{0}^{\dagger}a_{0}\right\rangle =1$ over the
measurement time%
\begin{equation}
S=\left\langle \int_{0}^{t_{\mathrm{m}}}d\tilde{Q}\right\rangle =2\sqrt
{2N_{1}+1}\sqrt{2k}t_{\mathrm{m}},
\end{equation}
with the noise $N$ over this averaging time%
\begin{equation}
N=\sqrt{\left\langle \int_{0}^{t_{\mathrm{m}}}d\tilde{Q}^{2}\right\rangle
}=\sqrt{2N_{1}+1}t_{\mathrm{m}}^{1/2}.
\end{equation}
Setting $S/N=1$ gives the measurement time%
\begin{equation}
t_{\mathrm{m}}=\frac{1}{8k}. \label{eq:collapsetime}%
\end{equation}
For jumps in the measured phonon number to be detected in the current we would
need $t_{\mathrm{dwell}}\gtrsim t_{\mathrm{m}}$. Notice that the measurement
time and the collapse time are comparable. This means that if the experimenter
can infer the system number state through the measurement current, then the
system is actually projected to that state on the same time scale. The results
for different averaging times $\Delta t$ is shown in Fig.\ \ref{fig:filter}.
For $k\Delta t$ equal to 4.5 or 7.5 the averaging is sufficient to display the
steps in $\left\langle a_{0}^{\dagger}a_{0}\right\rangle \left(  t\right)  $
without too much rounding of the transitions.

The simple averaging is not actually the optimal way to extract $\left\langle
a_{0}^{\dagger}a_{0}\right\rangle \left(  t\right)  $ from $I(t)$. In
principle a better approach is to use the stochastic master equation to
reconstruct the dynamics of the system given the initial ensemble
$\rho_{\mathrm{s}}(t_{0})$ and the measured current $I(t)$. This can be seen
more readily if we rewrite Eq.\ (\ref{eq:scaled-I(t)}) as%
\begin{equation}
dW=\frac{1}{\sqrt{2N_{1}+1}}\left\{  I\left(  t\right)  -2\sqrt{2k}%
\left\langle a_{0}^{\dagger}a_{0}\right\rangle \left(  t\right)  \right\}  dt.
\label{eq:dWexpression}%
\end{equation}
In our simulations we draw $I(t)$ at random from the appropriate distribution
and find the stochastic density matrix $\rho_{\mathrm{s}}$. However, using
$I(t)$ from experimental data, the experimenter can in principle propagate
Eq.\ (\ref{eq:numdist}) using Eq.\ (\ref{eq:dWexpression}) and then can
estimate the phonon number at each time from Eq.\ (\ref{eq:a0a0}). This
procedure is itself a low pass filtering that reduces the noise on the
measurement current, corresponding to the optimal filtering for our model of
the system\cite{b99,d99}

\section{Parameters and constraints}

\label{SubSec_parameter}

The results of the previous section show that a large value of the ratio
$k/\nu$ is crucial. To analyze the interplay of the parameters of an
experimental realization, we simplify the expression for $k/\nu$ by assuming
that most of the damping of the ancilla comes from the necessary coupling to
the measurement device, rather than from the extra thermal bath, \textit{i.e.}
$\mu\simeq\kappa$. Then using the expression of $k$ from Eq.\ (\ref{eq:k}) and
$\nu=\omega_{0}/2Q_{0}$ we get%
\begin{equation}
\frac{k}{\nu}\simeq4(2N_{1}+1)^{-1}Q_{0}Q_{1}\frac{\omega_{1}}{\omega_{0}%
}\left(  \frac{\lambda_{01}}{\omega_{1}}\right)  ^{2}|\alpha|^{2}.
\label{Eq:knu}%
\end{equation}

Equation (\ref{Eq:knu}) shows that the success of the measurement procedure is
favored by large oscillator quality factors, large driven response $\left\vert
\alpha\right\vert $, and a large value of the anharmonicity coupling factor
$\lambda_{01}/\omega_{1}$. In addition, as we have seen, detecting individual
jumps becomes harder as the temperature increases. Increasing the quality
factors of mesoscopic oscillators is an active area of research. Currently,
values of order $10^{3}~$to $10^{4}$ seem possible. If these could be raised
to the values characteristic of more macroscopic oscillators of the same
material, of order $10^{6}$ or even higher, the detection of individual
phonons would become correspondingly easier. The frequency ratio $\omega
_{1}/\omega_{0}$ appearing in Eq.\ (\ref{Eq:knu}) must be less than unity for
our detection scheme, but will probably not be too small because of geometry
constraints. Thus the main parameters available to optimize the experimental
geometry are the anharmonicity factor $\lambda_{01}/\omega_{1}$ and the
dimensionless measure of the driven displacement of the ancilla, $\left\vert
\alpha\right\vert ^{2}$ (the number of phonons in the driven state). We now
consider these factors in more detail.

\subsection{Anharmonicity Coefficient}

The interaction Hamiltonian for the system and ancilla oscillators
Eq.\ (\ref{eq:HRWA}) can be written%
\begin{equation}
H^{\mathrm{RWA}}=\hbar\omega_{0}a_{0}^{\dagger}a_{0}+\hbar\left[  \omega
_{1}+\lambda_{01}n_{0}\right]  a_{1}^{\dagger}a_{1},
\end{equation}
with $n_{0}$ the system phonon number. This equation implies that
$\lambda_{01}$ can be estimated as the frequency shift of the ancilla
oscillator for a single quantum ($n_{0}=1$) of the system oscillator.

For the prototype geometries using the two orthogonal flexing modes of a
single beam, or parallel flexing modes of two longitudinally coupled beams,
the nonlinear coupling arises from geometrical effects. At second order, the
transverse displacement in one mode gives a longitudinal strain, which then
changes the frequency of the second mode. The strain generated by the flexing
motion and the frequency shift associated with this strain can be derived
using elasticity theory, and have been calculated by Harrington and Roukes
\cite{HR94}. A demonstration of such frequency shift detection, and direct
measurement of $\lambda_{01}$\ between two coupled beam has recently been
reported\cite{HPYR03}. The longitudinal strain produced by a single quantum in
the fundamental flexing motion is%
\begin{equation}
\chi\simeq\frac{\hbar}{m_{0}\omega_{0}}\frac{1}{L_{0}^{2}}, \label{eq:strain}%
\end{equation}
where $m_{0}$ is the mass and $L_{0}$ is the length of the system beam. Then
the ancilla frequency shift caused by this strain is%
\begin{equation}
\lambda_{01}=\omega_{1}\frac{\zeta}{2\pi^{2}}\chi\frac{L_{1}^{2}}{d_{1}^{2}},
\label{eq:anharmonic}%
\end{equation}
where $\zeta$ is a geometric factor ($\zeta=3$ for clamped beam boundary
conditions) and $L_{1},d_{1}$ are the length and thickness of the ancilla
beam, respectively. Introducing a dimensionless quantity,%
\begin{equation}
R\equiv\frac{\hbar^{2}}{m_{1}d_{1}^{2}}\frac{1}{\hbar\omega_{1}},
\end{equation}
then the scaled coupling coefficient can be expressed as%
\begin{equation}
\frac{\lambda_{01}}{\omega_{1}}=\frac{\zeta}{2\pi^{2}}\frac{m_{1}\omega
_{1}L_{1}^{2}}{m_{0}\omega_{0}L_{0}^{2}}R. \label{eq:anharmonicityfactor}%
\end{equation}
Since the factor of the ratio of the two mode parameters will not be too large
or small, the most important quantity determining the anharmonicity factor,
which must not be too small for the success of the measurement scheme, is the
dimensionless ratio $R$. This will typically be a small number. The need for
small devices is seen from the scaling of this parameter with the dimensions.

\subsection{Driving strength}

The detection scheme we have considered is to measure the phase of the driven
response of the ancilla oscillator. Since the detection scheme is
magnetomotive, it is natural to consider the use of magnetic driving in
estimating the size of the displacement parameter $\left\vert \alpha
\right\vert $. For magnetic driving using a current $I_{\mathrm{drive}}$ in a
magnetic field $B$ the dimensionless displacement can be estimated as
\begin{equation}
\left\vert \alpha\right\vert =Q_{1}\frac{BI_{\mathrm{drive}}L_{1}d_{1}}%
{\sqrt{2}\hbar\omega_{1}}\sqrt{R}.
\end{equation}
Again the important role of $R$ in limiting the size of $\left\vert
\alpha\right\vert $ in this analysis is apparent.

We must also recognize that the size of $\left\vert \alpha\right\vert $ might
be limited by other physical constraints, rather than by the available drive
strength. One constraint might be to avoid undesired nonlinear effects in the
driven beam itself. For a classical oscillator, at sufficiently large drive
amplitudes nonlinear frequency pulling leads to a multiplicity of solutions
and instability. This occurs when the nonlinear frequency shift is comparable
with the width of the resonance $\omega_{1}/Q_{1}$. Using the same type of
estimate for the nonlinear frequency shift as in Eq.\ (\ref{eq:anharmonic})
shows that this occurs for%
\begin{equation}
\left\vert \alpha\right\vert \gtrsim\frac{1}{\sqrt{Q_{1}R}}.
\label{eq:nonlinearconstraint}%
\end{equation}
A more detailed, quantum mechanical analysis of the driven nonlinear
oscillator will be presented elsewhere \cite{qph04}.

\subsection{Example Configuration}

As a first estimate of the order of magnitude of the quantities introduced
above we will construct an example configuration using parameters that seem
plausible with current technology.

Recently, oscillators with resonant frequencies as high as $1$ GHz have been
fabricated \cite{HZMR03} using silicon carbide. Thus we consider two flexing
modes with resonant frequencies of $\omega_{0}=2.3$ GHz and $\omega_{1}=0.36$
GHz so that $\omega_{0}-\omega_{1}\gg\lambda_{01},\nu,\kappa$ are satisfied.
For this value of the system oscillator frequency, $\hbar\omega_{0}%
/k_{\mathrm{B}}T$ is unity at a temperature of about $0.1$ \textrm{K.} The
oscillators in Ref.\ \cite{HZMR03} were not very small, but it is expected
that the structure can be scaled down while maintaining the high oscillation
frequency. We therefore suppose smaller dimensions consistent with these
frequencies, namely dimensions are $0.6$ $\mu$m $\times$ $0.04$ $\mu$m
$\times$ $0.07$ $\mu$m for the system beam and $0.6$ $\mu$m $\times$ $0.04$
$\mu$m $\times$ $0.01$ $\mu$m for the ancilla beam. With these parameters we
obtain%
\begin{equation}
R=4.26\times10^{-9}. \label{eq:R}%
\end{equation}

The factor $R$ occurs squared in $k/\nu$ (via Eq.
(\ref{eq:anharmonicityfactor}), which is required to be large, and so this
small factor must be mitigated by the other quantities in Eq.
(\ref{eq:anharmonicityfactor}), i.e.\ large values of Q and a large driven
amplitude $\left\vert \alpha\right\vert $. Suppose the Q of the system
oscillator is $Q_{0}=10000$ and $Q_{1}=1000$. For the size of $\left\vert
\alpha\right\vert $ first consider the magnetic driving. A magnetic field
$B=10$ Tesla and $I_{\mathrm{drive}}=1$ $\mu$A can raise the driven response
to $\left\vert \alpha\right\vert \sim10^{5}$. To reach $k/\nu\sim1$ the
nonlinear coupling $\lambda_{01}/\omega_{1}$ that is required is then%
\begin{equation}
\lambda_{01}/\omega_{1}=4.9\times10^{-8}.
\end{equation}
With the given beam dimensions and the geometric nonlinearity, the anharmonic
coupling coefficient is actually $\lambda_{01}/\omega_{1}=1.3\times10^{-11}$,
about three orders of magnitude smaller. One possible way to increase this
value might be to engineer the geometry of the oscillator so that the
anharmonic coupling is larger than in the simple geometric nonlinearity we
have considered\cite{SP:R}. Another way to increase $k/\nu$ is to increase
$\left\vert \alpha\right\vert $ using a different driving scheme, although for
the value or $R$ in Eq.\ (\ref{eq:R}) the limit in Eq.
(\ref{eq:nonlinearconstraint}) is already exceeded for $\left\vert
\alpha\right\vert \gtrsim10^{3}$ , so that engineering the geometry to reduce
the self-nonlinearity might be necessary. An obvious way to increase $k/\nu$
to values greater than unity is to use oscillators with smaller dimensions,
for example carbon nanotubes.

\section{Conclusion}

We have analyzed a scheme to observe quantum transitions of a mesoscopic
mechanical oscillator. The non-linear coupling shifts the frequency of a
second (ancilla) oscillator proportionally to the excitation level of the
first (system) oscillator. This frequency shift may be detected as a phase
shift of the ancilla oscillation when driven on resonance. In principle, a QND
measurement is possible if the coupling constant between the two oscillators
$\lambda_{01}$ is much smaller than the resonance frequencies of the
oscillators, as will usually be the case. We have derived the master equation
for the system density matrix first integrating out the environment and
measurement degrees of freedom, and then by removing the ancilla operator
using the fact that the time scale of the system and ancilla dynamics are
quite different. The master equation has three components: phase diffusion as
a result of the measurement backaction; a constant energy shift due to the
excitation of the ancilla oscillator, and number state transitions due to the
interaction with the thermal bath (the environment).

The measurement process introduces a stochastic component into the system
dynamics and we have obtained the stochastic master equation corresponding to
our measurement scheme. From the stochastic master equation we identify two
competing tendencies that can be characterized by two parameters. One is the
coupling strength $\nu$ of the system and thermal bath, which is associated
with the dwell time $t_{\mathrm{dwell}}$\ between transitions. The other is
the coefficient $k$, associated with measurements, which includes not only the
coupling strength of the system to the measurement bath but also the
anharmonic coupling strength between the oscillators, the driving amplitude.
This coefficient is related to the measurement time $t_{\mathrm{m}}$\ that is
needed for a measurement to be able to produce an outcome with certainty. To
observe clear quantum jumps we would need $t_{\mathrm{dwell}}\gg
t_{\mathrm{m}}$. If this condition is not satisfied, then the experimenter
cannot infer the energy eigenstate of the system from the observed current.

Although our simple estimates based on plausible lithographically prepared
oscillators yield values for the ratio $t_{\mathrm{dwell}}/t_{\mathrm{m}}$ too
small for the observation of individual phonons, enhancements to the geometry
and the trend to smaller device sizes should improve the outlook. The basic
scheme and theoretical techniques developed here are fairly general, and in
particular are not restricted to zero temperature, and so can be also used for
other applications such as single spin detection and noise analysis for a
solid state based quantum computer. Such possibilities might open up a new
stage for observing quantum dynamics in mesoscopic systems.

\begin{acknowledgments}
We thank Michael Roukes for providing information about the experimental
progress in his group. DHS is grateful to Gerard Milburn and Tony Leggett for
insightful discussions and thanks for the Institute for Quantum Informaion at
Caltech for their hospitality. This work was supported by DARPA DSO/MOSAIC
through grant N00014-02-1-0602 and NSF through grant DMR-9873573. DHS's work
is also supported by the NSF through a grant for the Institute for Theoretical
Atomic, Molecular and Optical Physics at Harvard University and Smithsonian
Astrophysical Observatory. DHS also acknowledges the Weitzman fund for a
travel grant. Work by ACD was supported by the NSF through grant EIA-0086083
as part of the Institute for Quantum Information and the Caltech MURI Center
for Quantum Networks (DAAD19-00-1-0374).
\end{acknowledgments}

\appendix

\section{Bath field operators\label{Subsec:bathfield}}

In this appendix we describe more fully the time-local measurement bath
operators $B_{t}$ introduced in section \ref{SubSec_stochastic term}. The
description in terms of finely spaced modes of the bath with a smooth density
of states leads to the short memory or Markov property of the bath, that can
be expressed in terms of the time-local commutation rules for $B_{t}$. In the
main text we introduced the global bath operator as Eq.\ (\ref{eq:Bt})
\begin{equation}
B_{t}=\frac{1}{\sqrt{2\pi\rho_{\mathrm{d}}(\omega_{1})}g_{\mathrm{d}}\left(
\omega_{1}\right)  }\sum_{n}g_{\mathrm{d}}\left(  \omega_{n}\right)
b_{\mathrm{d},n}e^{-\iota\left(  \omega_{n}-\omega_{1}\right)  t}.
\label{eq:B(t)from bn}%
\end{equation}

We first derive the commutation rule Eq.\ (\ref{eq:Btcommutation}).
Substituting Eq.\ (\ref{eq:B(t)from bn}) in the commutator gives%
\begin{equation}
\left[  B\left(  t\right)  ,B^{\dagger}\left(  t^{\prime}\right)  \right]
=\frac{1}{2\pi\rho_{\mathrm{d}}\left(  \omega_{1}\right)  \left[
g_{\mathrm{d}}\left(  \omega_{1}\right)  \right]  ^{2}}\sum_{n,n^{\prime}%
}g_{\mathrm{d}}\left(  \omega_{n}\right)  g_{\mathrm{d}}\left(  \omega
_{n^{\prime}}\right)  \left[  b_{\mathrm{d},n},b_{\mathrm{d},n^{\prime}%
}^{\dagger}\right]  e^{-i\left(  \omega_{n}-\omega_{1}\right)  t}e^{-i\left(
\omega_{n^{\prime}}-\omega_{1}\right)  t^{\prime}}%
\end{equation}
and using $\left[  b_{\mathrm{d},n},b_{\mathrm{d},n^{\prime}}^{\dagger
}\right]  =\delta_{n,n^{\prime}}$ we obtain%
\begin{equation}
\left[  B\left(  t\right)  ,B^{\dagger}\left(  t^{\prime}\right)  \right]
=\frac{1}{2\pi\rho_{\mathrm{d}}\left(  \omega_{1}\right)  \left[
g_{\mathrm{d}}\left(  \omega_{1}\right)  \right]  ^{2}}\sum_{n}\left[
g_{\mathrm{d}}\left(  \omega_{n}\right)  \right]  ^{2}e^{-i\left(
\omega-\omega_{1}\right)  \left(  t-t^{\prime}\right)  }.
\end{equation}
Changing the sum to integral form%
\begin{equation}
\sum_{n}\rightarrow\int_{0}^{\infty}d\omega\,\rho_{\mathrm{d}}\left(
\omega\right)  ,
\end{equation}
gives%
\begin{equation}
\left[  B\left(  t\right)  ,B^{\dagger}\left(  t^{\prime}\right)  \right]
=\frac{1}{2\pi\rho_{\mathrm{d}}\left(  \omega_{1}\right)  \left[  g_{d}\left(
\omega_{1}\right)  \right]  ^{2}}\int_{0}^{\infty}d\omega\rho_{\mathrm{d}%
}\left(  \omega\right)  \left[  g_{d}\left(  \omega\right)  \right]
^{2}e^{-i\left(  \omega-\omega_{1}\right)  \left(  t-t^{\prime}\right)  }.
\end{equation}
Since $\rho_{\mathrm{d}}\left(  \omega\right)  $ and $g_{d}(\omega)$ are
slowly varying functions around the ancilla oscillation frequency
$\omega=\omega_{1}$ we can approximate these as$\ \rho_{\mathrm{d}}\left(
\omega\right)  \simeq\rho_{\mathrm{d}}\left(  \omega_{1}\right)  ,$
$g_{d}(\omega)$ $\simeq g_{d}(\omega_{1})$ and pull them outside of the
integral. Then introducing $\varepsilon=\omega-\omega_{1}$ and extending the
lower range of the integration over $\varepsilon$ to $-\infty$ leads to the
desired result%
\begin{equation}
\left[  B\left(  t\right)  ,B^{\dagger}\left(  t^{\prime}\right)  \right]
=\frac{1}{2\pi}\int_{-\infty}^{\infty}d\varepsilon e^{-i\varepsilon\left(
t-t^{\prime}\right)  }=\delta\left(  t-t^{\prime}\right)  .
\end{equation}

The interaction Hamiltonian for the ancilla oscillator and the measurement
bath is, from Eq.\ (\ref{eq:fullHamiltonian}),
\begin{equation}
H_{\mathrm{int}}^{\mathrm{I}}=i\hbar\sum_{n}g_{\mathrm{d}}\left(  \omega
_{n}\right)  \left[  b_{\mathrm{d},n}^{\dagger}(t)a_{1}(t)-b_{\mathrm{d}%
,n}(t)a_{1}^{\dagger}(t)\right]  , \label{eq:RWA-couple}%
\end{equation}
where we have moved to the interaction picture with $a_{1}(t)=a_{1}%
e^{-i\omega_{1}t}$\ and $b_{\mathrm{d},n}(t)=b_{\mathrm{d},n}e^{-i\omega_{n}%
t}$ the ancilla and bath operators in this picture, and $g_{\mathrm{d}}\left(
\omega_{n}\right)  $ is the coupling strength. The interaction Hamiltonian can
be written in terms of the bath operators $B_{t}$ as
\begin{equation}
H_{\mathrm{int}}^{\mathrm{I}}(t)=i\hbar\sqrt{2\mu}\left(  B_{t}^{\dagger}%
a_{1}-B_{t}a_{1}^{\dagger}\right)  , \label{eq:HintBt}%
\end{equation}
where the coefficient $\mu$ is
\begin{equation}
\mu\equiv\pi\varrho_{\mathrm{d}}\left(  \omega_{1}\right)  \left\vert
g_{\mathrm{d}}\left(  \omega_{1}\right)  \right\vert ^{2}%
\end{equation}
as before, and we have used the fact that the ancilla interacts predominantly
with bath modes near frequency $\omega_{1}$ and again have assumed a smooth
variation of the density of states and coupling constant so that we can make
the replacements $\rho_{\mathrm{d}}\left(  \omega\right)  \simeq
\rho_{\mathrm{d}}\left(  \omega_{1}\right)  $, and $g_{\mathrm{d}}(\omega)$
$\simeq g_{\mathrm{d}}(\omega_{1})$.

\end{document}